\documentclass{optica-article}

\journal{opticajournal} %

\articletype{Research Article}

\usepackage{lineno}
\usepackage{xcolor}

\begin{document}

\title{Designing lensless imaging systems to maximize information capture}

\author{Leyla~A.~Kabuli,\authormark{1,*} Henry~Pinkard,\authormark{1} Eric~Markley,\authormark{2} Clara~S.~Hung,\authormark{1} and Laura~Waller\authormark{1}}

\address{\authormark{1}Department of Electrical Engineering and Computer Sciences, University of California, Berkeley, California, USA\\
\authormark{2}
Graduate Program in Bioengineering, University of California, Berkeley and University of California, San Francisco, California, USA \\}
\email{\authormark{*}lakabuli@berkeley.edu} %

\begin{abstract*} 
Mask-based lensless imaging uses an optical encoder (e.g. a phase or amplitude mask) to capture measurements, then a computational decoding algorithm to reconstruct images. In this work, we evaluate and design lensless encoders based on the information content of their measurements using mutual information estimation. Our approach formalizes the object-dependent nature of lensless imaging and quantifies the interdependence between object sparsity, encoder multiplexing, and noise.
Our analysis reveals that optimal encoder designs should tailor encoder multiplexing to object sparsity for maximum information capture, and that all optimally-encoded measurements share the same level of sparsity. 
Using mutual information-based optimization, we design information-optimal encoders for compressive imaging of fixed object distributions. Our designs demonstrate improved downstream reconstruction performance for objects in the distribution, without requiring joint optimization with a specific reconstruction algorithm. 
We validate our approach experimentally by evaluating lensless imaging systems directly from captured measurements, without the need for image formation models, reconstruction algorithms, or ground truth data.
Our comprehensive analysis establishes design and engineering principles for lensless imaging systems, and offers a model for the study of general multiplexing systems, especially those with object-dependent performance.

\end{abstract*}
\section{Introduction}
Traditional optical design focuses on optimizing hardware to directly capture images of objects. In contrast, computational imaging systems expand system design into two parts: encoding measurements with non-traditional optics, and decoding measurements using a reconstruction algorithm. For example, mask-based lensless imagers encode images of objects using a thin amplitude or phase mask placed on top of an image sensor. The captured measurement looks nothing like the object, but a decoder is able to recover the object from the measurement by solving an inverse problem~\cite{lenslessreview}. Lensless imagers offer compact form factor and wide field-of-view, showing promise in applications including photography~\cite{diffusercam, phlatcam}, mesoscopy~\cite{biomesoscope, xueCM2}, and microscopy~\cite{diffuserscope, bioflatscope, flatscope}. 

While traditional imaging systems are designed to map each point of an object to a point on the sensor, lensless imaging systems instead map each object point to multiple points on the sensor, forming a multiplexed measurement. This multiplexing is necessary for compactness, since light cannot be concentrated fully when the distance between the mask and sensor is small. Multiplexing is also advantageous because it enables compressive encoding; for example, a single 2D measurement can be used to reconstruct 3D~\cite{diffusercam, phlatcam}, spatio-temporal~\cite{rollingshutterdiffusercam}, and hyperspectral~\cite{spectraldiffusercam} object properties. In 2D imaging, multiplexing enables  extended field-of-view imaging~\cite{YaokevinFOV, claraExtendedFOV, EllinNicoquadcam}, tolerance to sensor erasure~\cite{MonakhovaErasure}, and uniform resolution across a wide field-of-view~\cite{YangextendedSVAview, KCextendedSVAview}.

Optimal design of lensless imaging encoders is still an open problem. Most previously proposed designs are driven by heuristics; for example, phase masks are preferable to amplitude masks~\cite{flatcam, flatscope} for their light efficiency, with previous designs including off-the-shelf Gaussian diffusers~\cite{diffusercam}, contour masks that preserve good frequency transfer~\cite{phlatcam, bioflatscope, biomesoscope}, and lenslet arrays, which limit the amount of multiplexing to improve system performance~\cite{fenglenslet, xueCM2, diffuserscope, myrml, YaokevinFOV}. Fabrication advances now provide the flexibility to realize virtually any phase profile~\cite{kcphasemask, nanoscribefabex}. What is still needed are principled methods for determining optimal designs.

Typically, a candidate encoder design is evaluated based on the quality of the reconstructed images. Quality can be quantified by error metrics like peak signal-to-noise ratio (PSNR) or structural similarity index measure (SSIM); however, these require paired ground truth data. In the absence of ground truth data, quantifying image quality is challenging. Further confounding the evaluation of encoder design is the fact that the final image quality incorporates the combined effect of the encoder and decoder. Reconstruction quality depends heavily on the choice of decoder, priors, and hyperparameters~\cite{FISTA, ADMM, kristinalearning}, which makes it difficult to isolate the impact of the encoder.
For example, when reconstructing 3D data from a 2D measurement, sparsity constraints are crucial~\cite{candessparsity, diffusercam} and the performance impact of these constraints is object-dependent. Furthermore, as algorithms shift from physics-based models~\cite{kristinalearning} to deep neural networks~\cite{lenslesstransformer, convnext,kristinalearning} trained on large datasets~\cite{mirflickr, flatnetdataset, claradataset}, the state-of-the-art is continually updated. Ultimately, the performance of any of these inverse problem algorithms is limited by the quality of the encoded measurements. This highlights the need for a principled, decoder-independent evaluation method for lensless imaging.

Decoder dependence is a key limitation of end-to-end design, in which an encoder and decoder are jointly optimized with the goal of maximizing downstream performance (e.g. reconstruction quality)~\cite{originale2evincent, chrishdre2e, deepstorm3D}. 
In lensless imaging, end-to-end efforts have been applied to amplitude masks and task-specific objectives such as classification~\cite{cai2024lenslessface, Bezzamprogrammablemask, loene2e}, rather than general image reconstruction. Encoder performance remains confounded with decoder performance in end-to-end design, resulting in encoder designs that are specific to a particular decoder architecture and may not remain optimal as decoders are updated. In addition, this optimization is computationally intensive, which can make design intractable for high-dimensional problems, and is sensitive to initialization and often highly non-convex~\cite{wolfgange2e}.  %

Traditional system design metrics such as condition number~\cite{diffusercam} and modulation transfer function (MTF)~\cite{phlatcam, fenglenslet, xueCM2} are decoder-independent, but do not account for object properties or non-white noise. 
To better evaluate these effects in multiplexing systems, foundational work in coded aperture imaging defined multiplexing patterns for optimal object-independent performance~\cite{Gottesman89}, and subsequent studies 
analyzed multiplexing in conjunction with detection noise and signal priors~\cite{plenopticmultiplexing, illuminationmultiplexing} by evaluating measurement contrast~\cite{extendedfovsensor} and other data statistics~\cite{olliemultiplexing, olliemultiplexinghadamard}. Mutual information, which naturally combines object properties, such as sparsity, with imaging system properties, including signal-to-noise ratio (SNR) and bandwidth~\cite{shannon1949}, has the flexibility to capture the combined effects of the object, encoder, and noise~\cite{ashokcompressiveinfo, oldinfoimaging, neifeldTSI, ashoksimpleimaging}. With recent developments in practical estimation methods~\cite{Pinkard2024}, mutual information is particularly promising for decoder-independent system evaluation. 

In this work, we evaluate and design encoders for lensless imaging by directly quantifying the information content of measurements through mutual information estimation~\cite{Pinkard2024}. Our method estimates information from a dataset of measurements and a noise model, evaluating system performance without requiring image formation models or ground truth data. This measurement-based approach explicitly incorporates object dependence by evaluating encoders with respect to specified object distributions, and naturally extends to data-driven optimization of object-dependent encoder designs for general downstream performance. Although mutual information does not explicitly account for decoder effects and is therefore not a guaranteed predictor of downstream performance, empirical studies demonstrate consistent agreement between mutual information and general tasks such as reconstruction~\cite{Pinkard2024, javidiintimg2022, ashokcompressiveinfo}.

Using this method, we study tradeoffs between encoder multiplexing, object sparsity, and detection noise in lensless imagers. Multiplexing enables compressive encoding and compact form factor, but too much multiplexing can be detrimental to system performance, especially for dense objects. We find that the ideal amount of multiplexing should be tuned based on object sparsity. Our analysis further shows that all optimally-encoded measurements have the same measurement sparsity.
We then demonstrate the design of information-optimal encoders for compressive imaging with multiplexing levels optimized for different object distributions, and observe that these designs lead to improved reconstruction performance. Finally, we analyze experimental lensless imaging systems and confirm that dense natural images require encoders with less multiplexing.
Our contribution establishes engineering and design principles for lensless imaging, quantifying fundamental performance limits and enabling practical system improvements.

\section{Encoder Evaluation Framework} 
\label{sec:methods} 

\begin{figure*}[!t]
\centering
\includegraphics[width=\textwidth]{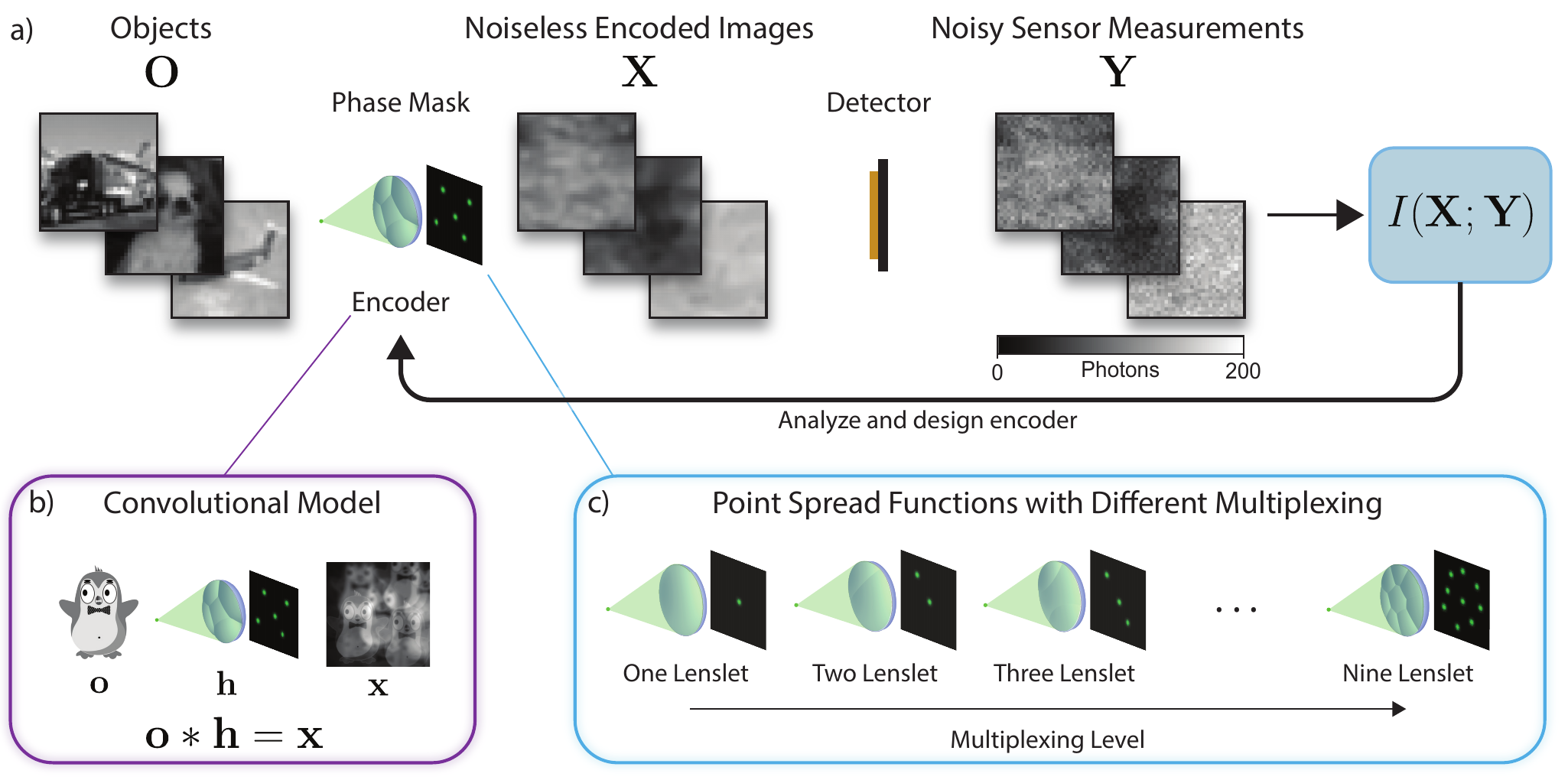}
\caption{Mutual information estimation framework for lensless imaging. a) A phase mask encoder maps an object distribution $\mathbf{O}$ to a noiseless image distribution $\mathbf{X}$. Image detection adds noise, forming a distribution of noisy sensor measurements $\mathbf{Y}$. Mutual information, estimated from the noisy sensor measurement distribution, serves as a metric for analyzing and designing the encoder. b) The lensless image formation model takes each object $\mathbf{o}$ and encodes it into a noiseless image $\mathbf{x}$ by convolving with the encoder point spread function $\mathbf{h}$. c) To analyze the effects of multiplexing in lensless imaging, we study encoders with one through nine lenslets, corresponding to increasing multiplexing levels.}
\label{fig:probmodel}
\end{figure*}

To study the interactions between objects and encoders, we estimate mutual information from measurements generated by various objects and encoders. This evaluation method is probabilistic, so it applies to a dataset of measurement samples rather than any individual measurement (Fig.~\ref{fig:probmodel}a).

First, the image formation process maps a distribution of objects $\mathbf{O}$ to a distribution of encoded images $\mathbf{X}$.
Specifically, each encoded image $\mathbf{x}$ is modeled as a convolution between the corresponding object $\mathbf{o}$ and a multiplexing point spread function (PSF) $\mathbf{h}$~\cite{diffusercam}
\begin{equation}
\label{eq:2Dconvolution}
    \mathbf{x} = \mathbf{o} \ast \mathbf{h}.
\end{equation}
The PSF is determined by the encoding phase mask (Fig.~\ref{fig:probmodel}b).
In practice, this image formation model also includes a crop due to the finite extent of the detector~\cite{diffusercam}. 

The goal of the encoder is to best capture object information to later be computationally recovered. While the objects and encoded images are distributions, there is only one encoder, which acts deterministically, such that repeated encodings of a given object produce the same encoded image. This encoding can be represented by a transformation $\mathbf{O} \rightarrow \mathbf{X}$ where the encoded image distribution $\mathbf{X}$ depends solely on the object distribution $\mathbf{O}$.

The encoded image distribution $\mathbf{X}$ is corrupted by noise at the sensor, forming the distribution of noisy sensor measurements $\mathbf{Y}$~(Fig.~\ref{fig:probmodel}a). We model the detection process with signal-dependent shot (Poisson) noise, which, in the high photon count regime, can be approximated as Gaussian with variance equal to the photon count at each pixel~\cite{goodmanstatisticaloptics}. The detection process does not depend on any knowledge of the specific encoder or objects being imaged; it simply introduces detection noise on the encoded measurements, represented by a transformation $\mathbf{X} \rightarrow \mathbf{Y}$.
The complete imaging model can be represented by a Markov chain $\mathbf{O} \rightarrow \mathbf{X} \rightarrow \mathbf{Y}$ and follows the widely-used semi-classical theory of photodetection~\cite{goodmanstatisticaloptics, ashoksimpleimaging}.

Our goal is to estimate the mutual information between objects and measurements, $I(\mathbf{O}; \mathbf{Y})$, which quantifies how well objects can be distinguished from each other based on the corresponding noisy measurements; essentially, how much object information survives the encoding and detection processes.
Although we may not have access to the true object distribution for estimating $I(\mathbf{O}; \mathbf{Y})$ directly, we can equivalently estimate $I(\mathbf{X}; \mathbf{Y})$, which measures how much of the object information captured in encoded images survives the detection process~\cite{Pinkard2024, neifeldTSI, ashoksimpleimaging}. Applying the Data Processing Inequality to our imaging model Markov chain, $I(\mathbf{O}; \mathbf{Y}) \leq I(\mathbf{X}; \mathbf{Y})$~\cite{dataprocessinginequality}, and since the encoding process is fixed with no additional sources of randomness, it follows that $I(\mathbf{O};\mathbf{Y}) = I(\mathbf{X}; \mathbf{Y})$ (Supplement Sec.~S1.1). 

Mutual information is decomposed as
\begin{equation}
    \label{eq:MIdecomp}
    I(\mathbf{X} ; \mathbf{Y}) = H(\mathbf{Y}) - H(\mathbf{Y} | \mathbf{X}).
\end{equation}
\noindent $H(\mathbf{Y})$ is the entropy of the noisy measurements, which captures variations across both the object and noise distributions, and $H(\mathbf{Y} | \mathbf{X})$ is the conditional entropy (randomness) due to noise alone. 
We adopt the mutual information estimation approach outlined in Pinkard et al.~\cite{Pinkard2024}, which consists of fitting a probabilistic model to the measurements and compensating for detection noise. 

The entropy $H(\mathbf{Y})$ quantifies the randomness in the noisy measurements. 
Entropy can be approximated by fitting a probabilistic model $p_\theta(\mathbf{y})$ to noisy measurement samples from the true measurement distribution $p(\mathbf{y})$ and calculating the cross-entropy $\mathbb{E}\left[ - \log p_\theta(\mathbf{Y})\right]$ on a held-out test set
\begin{equation}
    H(\mathbf{Y}) \leq \mathbb{E}\left[- \log p_\theta(\mathbf{Y})\right]
    \approx - \frac{1}{N} \sum_{i=1}^N \log p_\theta (\mathbf{y}^{(i)}),
    \label{eq:MIentropy}
\end{equation}
\noindent where $\mathbf{y}^{(i)}$ is the $i$th measurement in a test set of $N$ measurements~\cite{Pinkard2024, crossentropylaws}.

We choose a PixelCNN as the probabilistic model $p_\theta(\mathbf{y})$~\cite{pixelCNN} because it is powerful and flexible enough to accurately fit distributions of both dense and sparse images from limited samples~\cite{Pinkard2024}. Unless otherwise specified, we use 10,000 noisy measurement patches ($32 \times 32$ pixels) to fit the model and evaluate cross-entropy on a test set of 1,500 noisy measurement patches~\cite{traintestsplit}. Each model fit and evaluation takes approximately ten minutes and uses 3GB memory on an NVIDIA RTX A6000 GPU.

The conditional entropy $H(\mathbf{Y} | \mathbf{X})$ quantifies the randomness introduced by detection noise, which carries no information about the object.
We make the common assumption that the noise at each pixel in an $M$-pixel measurement is independent of other pixels, conditioned on the corresponding encoded image pixel $x_k$~\cite{goodmanstatisticaloptics}. For Poisson (shot) noise, this results in a closed-form expression for the conditional entropy,
\begin{equation}
    \label{eq:MIconditionalentropy}
    H(\mathbf{Y} | \mathbf{X}) \approx \frac{1}{N} \sum_{i = 1}^{N} \sum_{k=1}^{M} \frac{1}{2} \log_2  (2 \pi e x_k^{(i)}),
\end{equation}
\noindent which is a sum of conditional entropies across the $M$ image pixels averaged across a set of $N$ images drawn from the encoded image distribution~\cite{Pinkard2024}. This approximation is accurate at high photon counts ($>10$ photons), where shot noise can be approximated as Gaussian with equal mean and variance~\cite{goodmanstatisticaloptics, Pinkard2024}. Our final estimate of mutual information $I(\mathbf{X}; \mathbf{Y})$ is the difference between our upper bound on $H(\mathbf{Y})$ (Eq.~\eqref{eq:MIentropy}) and the value of $H(\mathbf{Y}| \mathbf{X})$ (Eq.~\eqref{eq:MIconditionalentropy}). We report mutual information estimates with 95\% confidence intervals from bootstrap estimates of cross-entropy and conditional entropy to quantify uncertainty. In Supplement Sec. S3.2 we also provide conditional entropy expressions and corresponding system analysis for Gaussian (read) noise.

\section{Quantifying Tradeoffs Between Sparsity and Multiplexing in 2D Imaging}
\label{sec:optimalsparsity2D}

\begin{figure*}[!t]
\centering
\includegraphics[width=0.55\textwidth]{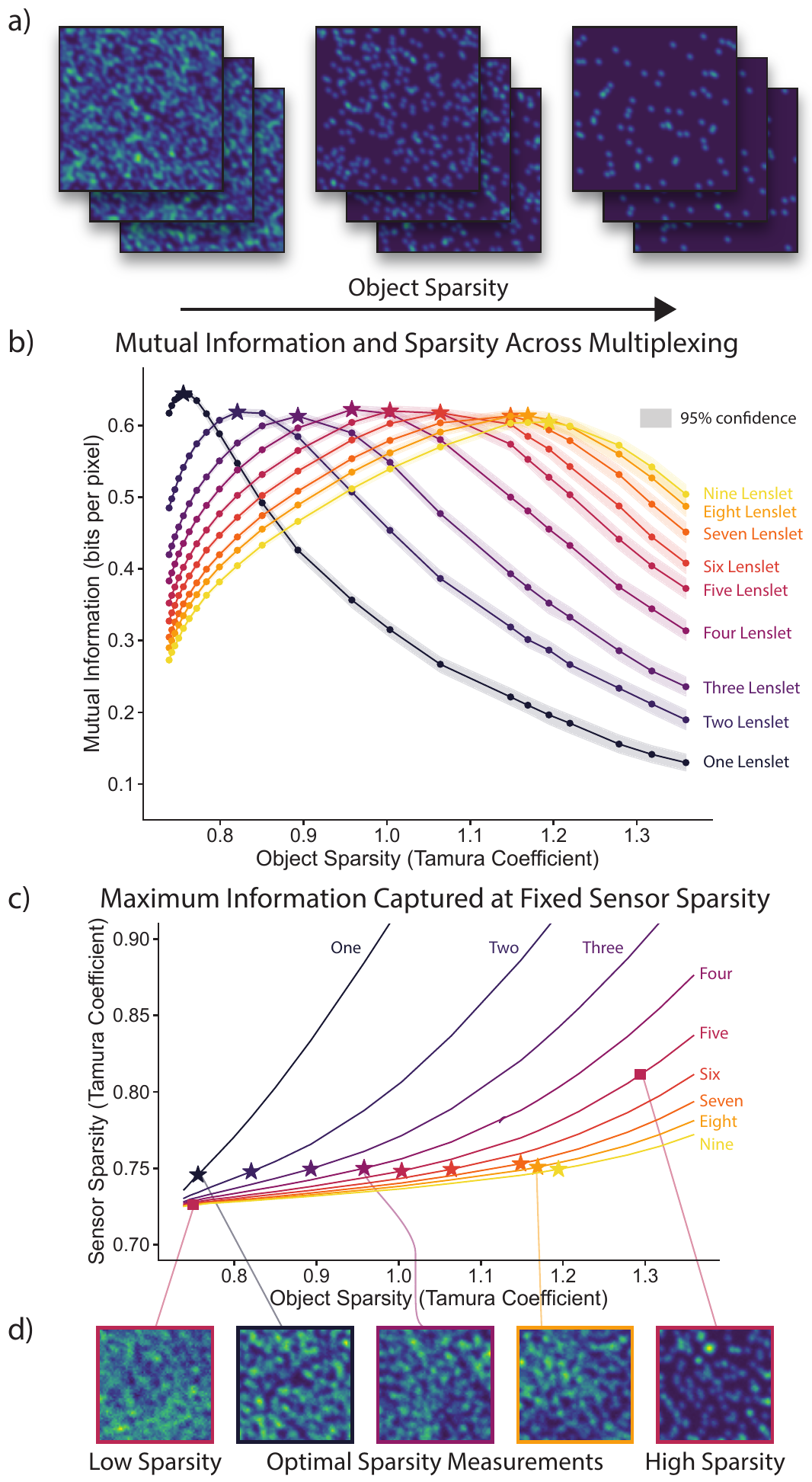}
\caption{Mutual information for varying object sparsity and encoder multiplexing. a) Examples of simulated objects with increasing levels of sparsity as quantified by the Tamura coefficient. b) Mutual information for one through nine lenslet multiplexing systems, each swept across object sparsity levels. As multiplexing increases, the maximum mutual information (denoted by stars) is achieved at higher sparsity levels. c) Maximum mutual information (denoted by stars) corresponds to a fixed sensor sparsity across all multiplexing encoders. d) Example measurements corresponding to optimal fixed sensor sparsity qualitatively differ from examples with low or high sparsity.}
\label{fig:beadimage2D}
\end{figure*}

Traditionally, imaging systems across microscopy, photography, and astronomy have been designed to have a single diffraction-limited spot as the PSF. These systems assume no computational decoding and minimize encoder multiplexing to form measurements that are directly human-interpretable. In contrast, lensless imagers are intentionally designed with some amount of encoder multiplexing, which enables compressive object encoding and recovery. This compressive imaging ability relies on object sparsity, making the performance of these multiplexing systems inherently object-dependent.  In order to provide design guidance, we aim to understand the fundamental tradeoffs between object sparsity and encoder multiplexing. To study these tradeoffs, we combine mutual information estimation with scalar metrics quantifying both sparsity and multiplexing.

Sparse objects are those that can be represented in a basis with few non-zero coefficients~\cite{candessparsity, sparsitymetriccriteria, candestaorecovery, donohocompressedsensing}. 
Sparsity is a critical assumption used in lensless imaging, as decoders harness object sparsity constraints to enable compressive recovery from captured measurements. For example, total variation (TV) sparsity can be used as a constraint when imaging objects with sparse gradients and edges, such as natural objects~\cite{totalvariation}. We quantify sparsity, specifically gradient sparsity, using the Tamura coefficient (TC)~\cite{Tamura1978}, a scalar metric previously used in holography~\cite{tamurasparsity, olderholotamura} and extended field-of-view imaging~\cite{tamurafov}. 

The Tamura coefficient of an image is 
\begin{equation}
\label{eq:tamuraeq}
TC(G) = \sqrt{\frac{\sigma_G}{\langle G \rangle}},
\end{equation}
\noindent where $G$ is the image gradient magnitude, $\sigma_G$ is the standard deviation of the gradient magnitude, and $\langle G \rangle$ is the mean of the gradient magnitude. The TC captures image gradient variability, where a higher TC corresponds to more sparsity and a lower TC corresponds to less sparsity. The high noise tolerance and sensitivity to signal changes~\cite{tamurasparsity} of the TC is suitable for our analysis of sparse objects and encoded noisy measurements. We utilize the TC to quantify the sparsity of object and measurement distributions by averaging TC values across samples in the corresponding distributions.

To quantify encoder multiplexing, we use the number of non-zero points in the encoder PSF. Here, we restrict our analysis to the class of lenslet-based encoders~\cite{diffuserscope, myrml, fenglenslet, xueCM2}, which have good SNR performance and provide a simple measure of multiplexing: the number of lenslets corresponds to the number of points in the PSF. We consider multiplexing levels ranging from a single lens (no multiplexing) through nine lenslets (high multiplexing) and show examples of heuristically-designed encoders and PSFs for each multiplexing level in Fig.~\ref{fig:probmodel}c. Each lenslet's focal spot is modeled as a Gaussian profile with isotropic aperture shape. For 2D imaging analysis, all lenslets have equal focal lengths. Each encoder maintains a constant photon budget with a fixed total aperture size and a PSF normalized to have unit energy. %

We systematically study the parameter space of object sparsity and encoder multiplexing by evaluating the encoded information in various combinations of objects and encoders. We simulate a class of objects for which sparsity and other variables can be precisely controlled: randomly-placed point sources, qualitatively similar to bead samples encountered in fluorescence microscopy~\cite{diffuserscope}. These datasets are generated by placing Gaussian-blurred point sources at randomly-selected pixel locations. The percentage of pixels selected for point sources controls the object sparsity while keeping other variables fixed. We generate datasets of beads with sparsity levels ranging from TC = 0.739 to TC = 1.359 (Fig.~\ref{fig:beadimage2D}a). Each dataset has a constant sparsity level and consists of 50,000 $96 \times 96$ pixel objects with a mean photon count of 100 photons per pixel. To probe encoder multiplexing levels, we consider encoders ranging from a single lens (no multiplexing) to nine lenslets (high multiplexing), as visualized in Fig.~\ref{fig:probmodel}c. Then, we simulate measurements for each multiplexing encoder and object sparsity level following the simulated imaging pipeline in Fig.~\ref{fig:probmodel}a. 
We compute mutual information from the resulting noisy measurements.

In Fig.~\ref{fig:beadimage2D}b, we study the tradeoffs between object sparsity and encoder multiplexing. Mutual information is calculated across object sparsity levels for each multiplexing encoder. For low-multiplexing encoders, mutual information is maximized (denoted by stars) for objects with low sparsity. As multiplexing levels increase, mutual information is maximized with increasingly sparse objects.

Furthermore, by examining only the starred points corresponding to maximum mutual information in Fig.~\ref{fig:beadimage2D}b, we find a surprising shared property of all optimally-encoded measurements: they have the same \textit{measurement} sparsity. Measurement sparsity is quantified by applying Eq.~\eqref{eq:tamuraeq} to the noisy measurements. To highlight this further, we plot the increasing relationship between object and sensor sparsity and denote the sensor sparsity that corresponded to maximum mutual information for each multiplexing encoder with a star (Fig.~\ref{fig:beadimage2D}c). Optimal measurements, achieved by maximizing information throughput, have a constant sensor sparsity (TC $\approx$ 0.75) (Fig.~\ref{fig:beadimage2D}d).
Our finding is supported by a related phenomenon in communication systems, where the optimal signals for achieving channel capacity produce the same distributions~\cite{flatmeasurementsinfotheory}.

We verify that our findings generalize to other encoder classes by demonstrating that diffuser-based encoders also achieve the same optimal sensor sparsity in Supplement Sec. S3.1. We also confirm that the same trends and optimal sensor sparsity finding hold for read noise-limited imaging systems in Supplement Sec. S3.2, and for different object normalizations in Supplement Sec. S3.3.

These findings are consistent with general intuition about multiplexing in imaging systems. Traditional single lens systems are designed to image natural objects, which have low sparsity. Multiplexing encoders, as used in lensless imaging, maximize information encoding with sparse objects, with optimal designs depending on the combination of object sparsity and encoder multiplexing. Our results also have an interpretation in terms of measurement discrimination. Increased multiplexing results in measurements with overlapping object copies. For objects that are dense, these overlapping copies lead to measurement saturation, making measurements difficult to discriminate and reducing information. For extremely sparse objects, overlapping copies can effectively utilize regions of the sensor that would otherwise not detect signal. This redundancy and resulting noise variation can provide an advantage for discrimination, effectively increasing information. 

From a design perspective, these results indicate that the measurement sparsity generated by an encoder for a given distribution of objects can offer simple guidance for evaluating and adjusting the encoder.
For maximum information encoding, our findings suggest that using phase masks with more multiplexing can be beneficial when imaging sparse objects, and using phase masks with less multiplexing is preferable when imaging dense objects.

\section{Information-Optimal Encoder Design}
\label{sec:ideal}

\begin{figure*}[!t]
\centering
\includegraphics[width=\textwidth]{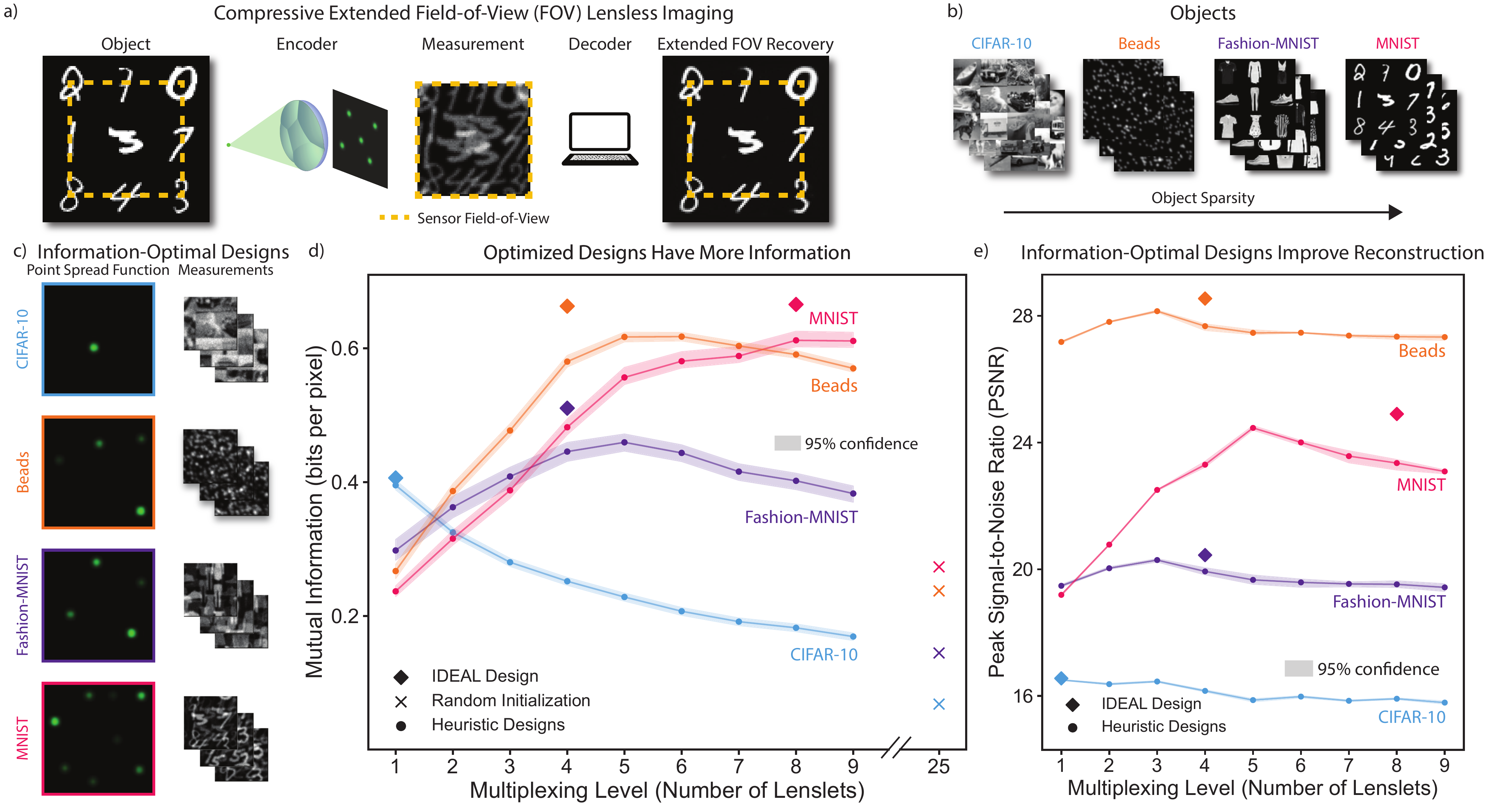}
\caption{Information-optimal phase mask design for lenslet-based encoders with Information-Driven Encoder Analysis Learning (IDEAL). 
a) Lensless imaging enables compressive extended field-of-view imaging in 2D by encoding an object larger than the finite extent of an image sensor into a limited measurement region through multiplexing. A decoder recovers an extended field-of-view from the measurement.
b) Examples of objects with varying sparsity.
c) Visualization of the information-optimal encoder, point spread function, and example measurements for each object distribution. As the object sparsity increases, the number of lenslets (multiplexing level) in the corresponding information-optimal design increases.
d) Mutual information for IDEAL designs for each object distribution. Random initialization with 25 lenslets is denoted by an "X" and the resulting IDEAL design is denoted by a diamond. Information-optimal designs have higher information than baseline heuristic designs (denoted by circles).
95\% confidence intervals are smaller than the marker size for IDEAL designs and indicated by shading for heuristic designs.
 e) Higher mutual information in IDEAL designs results in improved reconstruction performance for each object distribution, as quantified by reconstruction peak signal-to-noise ratio.}
\label{fig:IDEAL}
\end{figure*}

We studied a series of heuristically-designed encoders and identified the object sparsity levels best matched to each encoder in Sec.~\ref{sec:optimalsparsity2D}. Next, we explore the design of information-optimal encoders for lensless imaging that maximize encoded information for fixed object distributions. %

Here, we design multiplexing encoders for single-shot extended field-of-view imaging~\cite{YaokevinFOV, claraExtendedFOV}, in which our objective is to recover a 2D object region that is larger than the finite sensor extent (measurement), as visualized in Fig.~\ref{fig:IDEAL}a. 
We can use mutual information to design encoders with Information-Driven Encoder Analysis Learning (IDEAL), which has been used to design color filter arrays and pupil-plane masks with improved downstream reconstruction performance~\cite{ericideal, Pinkard2024, idealio}. IDEAL does not require a decoder. Instead, measurements from the imaging system are evaluated with mutual information as the loss and the encoder is optimized for a specific object distribution using gradient-based methods via backpropagation. This decoder-independent approach provides significant compute time and memory savings compared to end-to-end design~\cite{idealio} and eliminates the need for retraining for different reconstruction architectures, as algorithms can be flexibly swapped in afterwards.

To determine the information-optimal encoder for a specific object distribution $\mathbf{O}$, the design objective is defined by
\begin{equation}
    \mathbf{h}^{\ast}
=
\underset{\mathbf{h} \ge 0, \, \|\mathbf{h}\|_{1} = 1}{\arg\max}
\; I\!\left(f_{\mathbf{h}}(\mathbf{O})\,;\, \mathbf{Y}\right).
\label{eq:idealobjective}
\end{equation}
For a non-negative unit energy PSF $\mathbf{h}$,  the mapping $f_\mathbf{h}(\mathbf{O}) = \mathbf{X}$ applies a convolution with $\mathbf{h}$ to each object in $\mathbf{O}$ following the image formation model in Eq.~\eqref{eq:2Dconvolution}, and $\mathbf{Y}$ is the resulting noisy measurement distribution. 
At each optimization step, the encoder PSF is updated, the measurement distribution is recomputed based on this PSF and the fixed object distribution, and the mutual information loss is evaluated on these updated measurements. The final PSF $\mathbf{h}^{\ast}$ is the information-optimal design for the specified object distribution.

Solving the design objective in Eq.~\eqref{eq:idealobjective} with gradient-based optimization requires a differentiable model of the optical encoder. We extend the class of lenslet-based encoders studied in this work to a differentiable model by representing the PSF of the lenslet array as a sum of Gaussians. Each lenslet's point response is parameterized as a Gaussian with learnable parameters: mean (position), covariance (aperture shape), and weight (aperture size and relative energy). We constrain covariance matrices to be isotropic (round apertures), as designs with isotropic apertures achieved higher information than designs with anisotropic apertures (Supplement Sec. S5).

We consider the design of optimal encoders for four different object distributions: CIFAR-10~\cite{CIFAR10}, representing dense natural objects (TC = 0.970), Beads, representing moderately sparse objects (TC = 1.1064), Fashion-MNIST~\cite{Fashion-MNIST}, representing intermediate sparsity objects (TC = 1.196), and MNIST~\cite{MNIST}, representing sparse, binary objects (TC = 1.363). Each dataset is normalized to a mean photon count of 100 photons per pixel and object samples are visualized in Fig.~\ref{fig:IDEAL}b.

We use a PixelCNN to compute the mutual information loss~\cite{idealio}. This estimator is slower than the Gaussian estimation model previously used for IDEAL~\cite{Pinkard2024}, but has the flexibility to fit both dense and sparse distributions. At each iteration, measurements are generated from the updated encoder and the object distribution based on the image formation model in Fig.~\ref{fig:probmodel}a. The mutual information loss is calculated with 4096 $16 \times 16$ pixel measurement patches, of which 90\% are used to fit the PixelCNN, and the other 10\% for the evaluation test set. Based on the loss, lenslet parameters are updated to refine the encoder design, with the objective of maximizing the mutual information. The PixelCNN is reinitialized and refit every 20 encoder optimization steps to reduce computational runtime.
Each encoder optimization takes less than two hours and uses 5GB memory on an NVIDIA RTX A6000 GPU.

We evaluate baseline performance on heuristic lenslet designs for each object distribution and plot the information in Fig.~\ref{fig:IDEAL}d. The general trend is consistent with previous findings: dense objects benefit from lower multiplexing and sparse objects benefit from higher multiplexing. We then initialize our IDEAL optimization process with a random distribution of 25 lenslets (Fig. S7) and train to convergence for each object distribution. The mutual information for this random initialization encoder is denoted by an "X" for each dataset. We optimize the encoder by updating lenslet positions, widths, and relative weightings using IDEAL. We use a complete model fit with 10,000 $32\times32$ pixel patches and 1,500 patch test set to calculate mutual information for the resulting information-optimal design, denoted by a diamond in Fig.~\ref{fig:IDEAL}d. The information-optimal design for each object distribution has higher mutual information than any of the heuristic designs, with significant improvements on the different designs for the Beads, Fashion-MNIST, and MNIST datasets.

Figure~\ref{fig:IDEAL}c visualizes the resulting encoder PSF from IDEAL and an example measurement for each object distribution. The optimal design for encoding CIFAR-10 natural images is a single lens, which is expected because dense objects do not benefit from multiplexing. As a result, the information gain for the CIFAR-10 design is minimal, as there are limited degrees of freedom for designing a single lens. For both the Beads and Fashion-MNIST datasets, the optimal design converges to four lenslets, but with different weights and lenslet positions. For sparse MNIST objects, the optimal design converges to eight lenslets.

We verify downstream reconstruction performance for our information-optimal designs. For each object distribution and corresponding design, we reconstruct images with extended field-of-view from multiplexed measurements using a lightweight U-net (Supplement Sec.~S1)~\cite{kristinalearning}. Example reconstructions are visualized in Fig.~S7. In Fig.~\ref{fig:IDEAL}e, we compare average reconstruction performance (PSNR) for each IDEAL design and all heuristic designs. Although these information-optimal designs are not optimized for a specific decoder, the resulting measurements lead to improved reconstruction performance for the specified object distribution, demonstrating for this compressive lensless imaging task that increased information benefits downstream recovery using a U-net.

Optical encoder design is often non-convex and sensitive to initialization~\cite{wolfgange2e}. In Supplement Sec. S5, we verify our results from Fig.~\ref{fig:IDEAL}d by optimizing with 10 different random initializations, all of which converge to similar encoders that outperform heuristic designs. We also consider heuristic initialization, in which we optimize the positions of the lenslets while keeping other parameters, including multiplexing level, fixed. For sparse datasets including Beads and MNIST, designs with heuristic initialization achieve slightly higher information than with random initialization, and additionally produce measurements achieving optimal sensor sparsity (Sec.~\ref{sec:optimalsparsity2D}). Evaluating sensor sparsity can help determine whether a design is close to the optimum.

Our information-optimal designs for various objects verify our previous findings, including that sparse objects benefit from higher multiplexing (Sec.~\ref{sec:optimalsparsity2D}).
Future work can investigate manufacturing information-optimal phase masks for lensless imaging systems. Constraining optimization to equally-weighted lenslets can be simpler for manufacturing, although recent developments in lithography techniques have demonstrated relative ease of manufacturing of non-traditional lenslets and other phase masks~\cite{kcphasemask, nanoscribefabex}.

\section{Experimental Information Evaluation}
\label{sec:experimental_MI}

We demonstrated lensless imaging encoder evaluation and design in Sec.~\ref{sec:optimalsparsity2D} and Sec.~\ref{sec:ideal} via simulations. Now, we apply our approach to experimental data. Experimental measurements include additional imaging non-idealities including a shift-varying PSF. Mutual information is well-suited for experimental analysis, as it directly evaluates measurements without requiring an explicit image formation model. This captures the effects of non-idealities without the need for extensive system characterization or calibration (e.g. PSF capture).

\begin{figure*}[!t]
\centering
\includegraphics[width=0.65\textwidth]{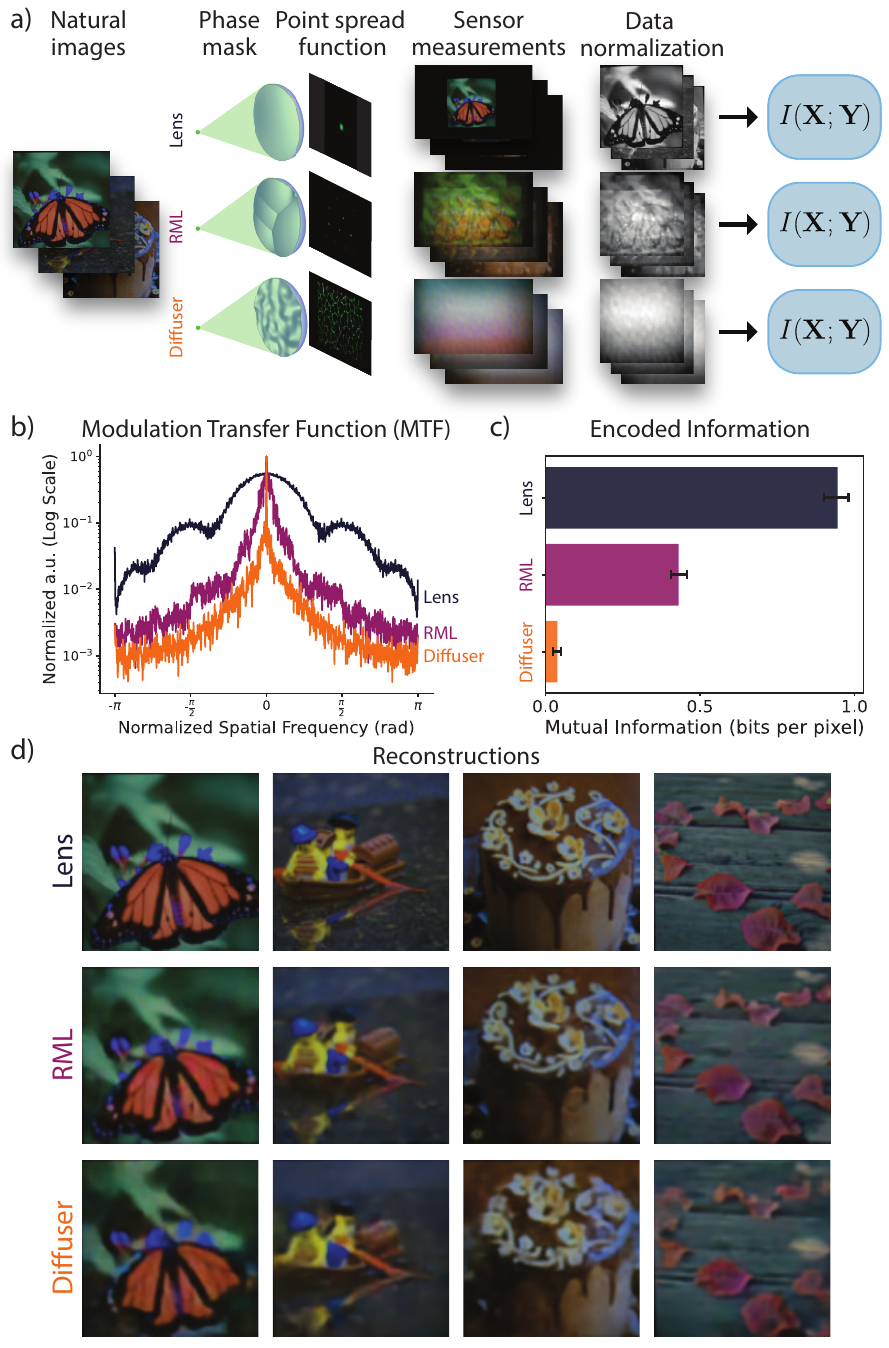}
\caption{Mutual information in experimental lensless imaging comparing a traditional lens to two phase mask encoders: a diffuser and a random multi-focal lenslet (RML) array. a) Mutual information is estimated from experimentally-captured measurements of natural images for each imaging system. b) 1D cross-section of the modulation transfer function for each encoder. The high-multiplexing diffuser has the worst frequency transmittance. c) The lens system has the most mutual information and information decreases with multiplexing. d) Examples of reconstructions for the RML and diffuser systems compared to the lens system show worse image quality with the high-multiplexing diffuser. The reconstruction structural similarity index measure (SSIM) for the RML (SSIM = 0.886) is higher than the diffuser (SSIM = 0.821).}
\label{fig:experimental}
\end{figure*}

We consider three encoders: a traditional lens with no multiplexing, a random multi-focal lenslet (RML) array phase mask~\cite{myrml} with moderate multiplexing, and a Gaussian diffuser phase mask~\cite{diffusercam} with high multiplexing. The object distribution is natural images represented by the MIRFLICKR-25000 dataset~\cite{mirflickr} (mean TC = 1.238, standard deviation 0.213) displayed on a monitor placed in the imaging plane. This data is from an open-access dataset consisting of measurements from each system captured in parallel under identical imaging conditions~\cite{claradataset}. 

To calculate mutual information, we calibrated the noise model and estimated the amount of shot noise. We converted raw camera data from pixel intensities to photon counts, assuming that shot noise is the dominant source of noise (equal variance and mean)~\cite{janesick2007}. We processed measurements by converting to grayscale, cropping to square dimensions, and downsizing to $100\times100$ pixel patches with pixel binning (Fig.~\ref{fig:experimental}a). To correct quantization artifacts from limited sensor bit depth, we added additional shot noise to simulate lower signal-to-noise ratio measurements.

Mutual information is estimated by fitting the PixelCNN model with 10,000 measurement patches, with a test set of $1,000$ patches for evaluation. Since experimental noisy measurements do not provide access to the noiseless image distribution $\mathbf{X}$, we use $\mathbf{Y}$ in place of $\mathbf{X}$ for conditional entropy estimation (Eq.~\eqref{eq:MIconditionalentropy}). This approximation is accurate when images have reasonably high photon count  ($> 10$ photons)~\cite{Pinkard2024}.

First, we evaluate system performance with decoder-independent evaluation methods, using mutual information and the MTF, which quantifies the filtering characteristics of the encoder in isolation without accounting for object or noise properties. In Fig.~\ref{fig:experimental}, we visualize each system's PSF and a 1D cross-section of the MTF, with 2D MTFs shown in Fig.~S6. As multiplexing increases, the MTFs have worse frequency transmittance (Fig.~\ref{fig:experimental}b). Mutual information estimation shows that information decreases as multiplexing increases (Fig.~\ref{fig:experimental}c). This is expected based on Sec.~\ref{sec:optimalsparsity2D}; dense natural images do not benefit from multiplexing. MTF-based analysis requires system PSF calibration and captures encoder properties in isolation, while mutual information does not require PSF calibration and additionally captures both object and noise properties.

We also compare direct measurement-based evaluation using mutual information with traditional reconstruction-based evaluation. For each lensless imager, we reconstruct images using a ConvNeXt architecture~\cite{convnext, convnext2020s} with a train set of 24,000 measurements and a test set of 1,000 measurements. All measurements are $8\times$ downsampled and the images from the lens system serve as ground truth. Reconstructions are visualized in Fig.~\ref{fig:experimental}d. The RML reconstructions achieve a test set structural similarity index measure (SSIM) of 0.886, while the diffuser reconstructions have worse SSIM of 0.821. The reconstruction quality is also qualitatively worse with more multiplexing, as fine details and sharp edges are missing in the diffuser reconstructions but recovered in the RML reconstructions. The decoder-independent information estimates are in agreement with reconstruction quality while avoiding the need for paired ground truth, which is often unavailable in experimental settings. This highlights a valuable aspect of our decoder-independent approach, as it enables quantitative evaluation directly from raw measurements.

Although our experimental results confirm that a traditional lens outperforms lensless systems for 2D natural images, lensless imagers are motivated by factors beyond image quality. These multiplexing systems enable compressive image encoding (e.g. extended field-of-view or depth), and can achieve ultra-thin form factor~\cite{flatscope}, capabilities not possible with traditional lenses. The RML and diffuser illustrate these tradeoffs: each sacrifices performance relative to a lens, as quantified with our mutual information analysis, while providing application-specific advantages in encoding capability and compactness.

\section{Conclusion}

Our contribution bridges a crucial gap in lensless imaging system evaluation by providing a comprehensive, measurement-based and decoder-independent analysis that captures the joint effects of object sparsity, encoder multiplexing, and detection noise. Our analysis revealed that optimal encoder designs should adjust encoder multiplexing to object sparsity for maximum information encoding. We demonstrated the design of object distribution-specific information-optimal encoders for compressive imaging across object sparsities and evaluated experimental measurements to verify the benefits of low multiplexing for dense objects. %
Together, these results highlight a shift toward encoder designs tailored to specific object distributions rather than general objects.

As lensless imaging systems advance towards applications including in vivo imaging~\cite{biomesoscope, bioflatscope}, designing encoders that optimally match properties of the target object distribution (e.g. sparsity) is crucial in order to facilitate high-quality imaging. Our framework can be used to discover new object-dependent encoder designs and to optimize the performance of existing empirical designs~\cite{xueCM2, myrml, YaokevinFOV}. Furthermore, our work serves as an example of how mutual information estimation can reveal design principles for multiplexing systems, paving the way for improving other computational imaging systems, especially those with object-dependent performance.

Future work can expand on our modeling of lensless imaging systems.
We primarily analyzed lenslet-based and diffuser-based (Supplement Sec. S3) encoders, revealing fundamental tradeoffs through simple image formation models. Other classes of phase masks and optical encoders can be designed and evaluated with the same approach. Simulation models can be extended to incorporate system-specific image formation constraints in order to probe tradeoffs including form factor or compressive sensing ability. For example, in Supplement Sec. S2 we analyze fundamental properties of compressive 3D imaging to demonstrate that depth multiplexing increases information, especially for sparse objects.
Finally, our detection models focused on signal-dependent shot noise with a complementary example for signal-independent read noise (Supplement Sec. S3). Incorporating imager-specific joint or learned noise models could offer additional insight into the effects of noise.

Our design principles for information-optimal phase mask design demonstrate how tuning encoder multiplexing to specific object sparsities can maximize performance for both heuristic and optimized designs. Our distribution-based approach is most effective when object properties are known or can be accurately modeled. The resulting insights and design choices are object distribution-specific rather than universally optimal, and may not generalize if the assumed object distribution is mismatched or if mutual information estimates are sample-limited.

Although mutual information provides a general measure of object-dependent system performance, specialized downstream tasks, such as classification, often rely on a subset of this information. Maximizing total mutual information therefore does not guarantee an increase in task-specific information or performance. To optimize imaging systems for specialized tasks rather than general performance, it may be beneficial to condition the information estimator on task-relevant components and only maximize task-specific information~\cite{Pinkard2024, ashokcompressiveinfo}. Similarly, reconstruction algorithms with limited flexibility or power may not be able to utilize all encoded information. Algorithm-specific conditioning, potentially combined with end-to-end system fine-tuning, could help identify designs that maximize the information most effectively used by a given reconstruction algorithm. Task-specific and algorithm-specific conditioning offer promising extensions for lensless imaging and mutual information-based design.

\begin{backmatter}
\bmsection{Funding}
\noindent
L.A.K. was supported by the National Science Foundation Graduate Research Fellowship Program under Grant DGE 2146752. L.W. is a Chan Zuckerberg Biohub SF investigator. This work was supported by the U.S. Air Force Office Multidisciplinary University Research Initiative (MURI) program under Award No. FA9550-23-1-0281, the U.S. Air Force Office of Scientific Research under Award No. FA955-22-1-0521, and STROBE: A National Science Foundation Science \& Technology Center under Grant No. DMR 1548924.

\bmsection{Acknowledgment}
The authors would like to thank Kannan Ramchandran, Vasilisa Ponomarenko, Amit Kohli, Tiffany Chien, and Emma Alexander for helpful discussions. L.A.K. was supported by the National Science Foundation Graduate Research Fellowship Program under Grant DGE 2146752. L.W. is a Chan Zuckerberg Biohub SF investigator.

\bmsection{Disclosures}
The authors declare no conflicts of interest.

\bmsection{Data Availability} Code and data presented in this paper are available in~\cite{ourcodegithub}.

\bmsection{Supplemental document}
See Supplementary Material for additional results and implementation details.

\end{backmatter}

\bibliography{references}

\newpage 

\renewcommand{\thesection}{S\arabic{section}}
\renewcommand{\thetable}{S\arabic{table}}
\renewcommand{\thefigure}{S\arabic{figure}}
\renewcommand\theequation{S\arabic{equation}}
\setcounter{figure}{0}
\setcounter{table}{0}
\setcounter{section}{0}
\setcounter{equation}{0}

\noindent
\textbf{\large Designing lensless imaging systems to maximize information capture: \\ Supplementary material}

\noindent 
This supplementary material contains additional results and implementation details.

\section{Implementation details}
\label{supp:implementation}

As described in the main text, we follow the mutual information estimation approach outlined in Pinkard et al.~\cite{Pinkard2024} to upper bound mutual information. To verify estimation consistency and ensure the tightest upper bound, we repeat all estimates with 5 different random trials (seeds). Reported information estimates are the minimum estimate across trials and the corresponding 95\% confidence interval. Information estimates across trials are consistent, with less than 0.01 bits per pixel range.

All simulated 2D objects are $96 \times 96$ pixels and all 2D PSFs are $32 \times 32$ pixels, resulting in $65 \times 65$ pixel measurements using valid mode convolution. This model represents compressive extended field-of-view lensless imaging, in which the measurement size is smaller than the object being imaged. For pre-existing datasets (CIFAR-10, Fashion-MNIST and MNIST), we randomly tile objects in a $3 \times 3$ grid to match these dimensions, including zero-padding on Fashion-MNIST and MNIST datasets to match the dimensions of CIFAR-10. This random tiling also prevents intensity falloff at the edges of the measurement field-of-view, which enables isolating the effects of object sparsity and encoder multiplexing on information without the confounding effect of finite object extent. All measurements include a 10 photon bias to ensure measurements are in the high photon count regime where the Gaussian approximation for shot noise is valid. We randomly select $32 \times 32$ pixel patches from across the measurement field-of-view for mutual information estimation. The PixelCNN model fit uses $1 \times 10^{-3}$ learning rate, 500 steps per epoch, and 20 epoch patience for early stopping. 

For information-optimal design, we use 2000 optimization steps. Mutual information estimates during optimization use a coarse PixelCNN model fit, with $16 \times 16$ pixel patches, $1 \times 10^{-3}$ learning rate, 100 steps per epoch, and 10 epoch patience. The PixelCNN is reinitialized and refit every 20 optimization steps. The tuned learning rates for stable convergence for the differentiable lenslet model are $3 \times 10^{-2}$ for the means, $1 \times 10^{-3}$ for the covariances, and $1 \times 10^{-4}$ for the weights. MNIST designs use $5 \times 10^{-3}$ learning rate for the means, and CIFAR-10 designs use $3 \times 10^{-4}$ for the weights.

\begin{figure}[!t]
\centering
\includegraphics[width=0.55\columnwidth]{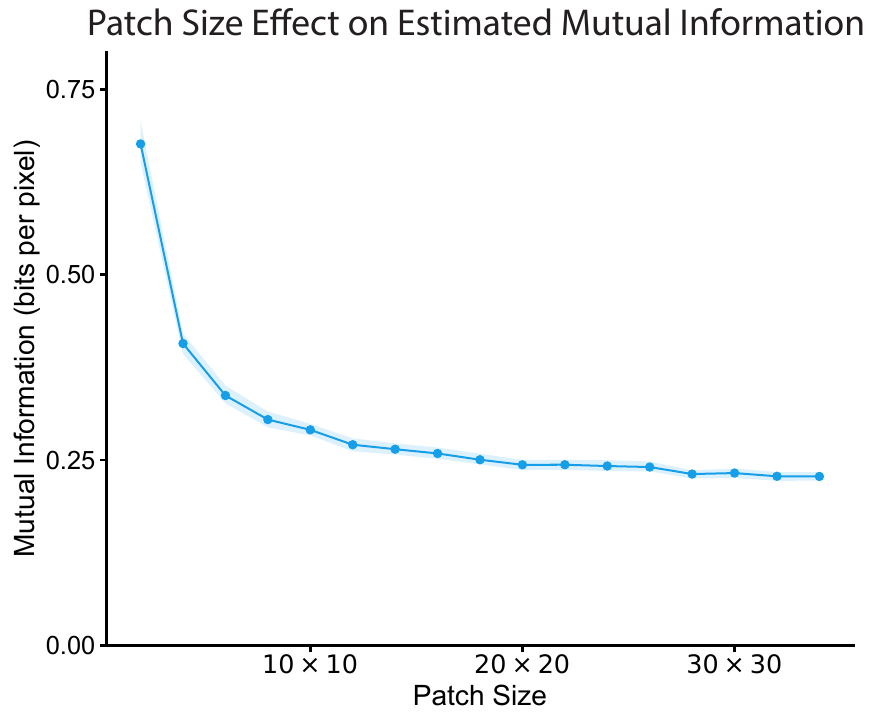}
\caption{The effect of patch size on estimated mutual information. Larger patch sizes result in lower estimated mutual information, and level off after $10 \times 10$ pixel patch size.}
\label{fig:patchsize}
\end{figure}

We study the effect of patch size on estimated mutual information in Fig.~\ref{fig:patchsize}. As the patch size increases, longer-range dependencies between pixels can be captured in patches, resulting in a tighter upper bound on mutual information. For patch sizes larger than $10 \times 10$, the mutual information values level off. The mutual information estimates have only 0.04 bits per pixel difference between $12 \times 12$ pixel patches and $32 \times 32$ pixel patches.

To reconstruct extended field-of-view objects from measurements (Fig.~3, Fig.~\ref{fig:initialization}, Fig.~\ref{fig:supplementIDEAL}), we adapt a lightweight version of a U-net used for lensless image reconstruction~\cite{kristinalearning}. This U-net uses two encoder-decoder stages to reduce complexity and runtime, which we find sufficient to utilize the information in measurements. It consists of an encoder with two convolutional blocks (24 and 64 channels), a 64-channel bottleneck, a decoder with two upsampling blocks, and an output layer mapping 24 channels to a single-channel output, following standard block structure~\cite{kristinalearning}.

The network takes as input $65\times 65$ pixel measurements with shot noise, zero-padded to $96\times 96$ and normalized to unit maximum, and outputs $96\times96$ pixel reconstructions. We train on 20,000 measurements, validate on 2,000, and test on 2,000, with a batch size of 64. Training uses mean squared error loss, the AdamW optimizer (learning rate $2 \times 10^{-3}$, weight decay $10^{-4}$, and a CosineAnnealing schedule), with early stopping (patience 10 epochs) over a maximum of 200 epochs. For each encoder, we train 5 networks to account for random variation and report mean test set PSNR with 95\% confidence intervals. Training each network takes approximately 10 minutes and 2GB memory on an NVIDIA RTX A6000 GPU. Example extended field-of-view reconstructions from information-optimal and heuristic lenslet designs are visualized in Fig.~\ref{fig:initialization}c and Fig.~\ref{fig:initialization}d.

\subsection{Proof of $I(\mathbf{O}; \mathbf{Y}) = I(\mathbf{X}; \mathbf{Y})$}
\label{sec:dpiequalityproof}

We represent our imaging system as a Markov chain $\mathbf{O} \rightarrow \mathbf{X} \rightarrow \mathbf{Y}$. $\mathbf{X} = f(\mathbf{O})$, where $f$ is a deterministic but not necessarily invertible function. $\mathbf{X} \rightarrow \mathbf{Y}$ depends only on the encoded images $\mathbf{X}$. 
We now show $I(\mathbf{O}; \mathbf{Y}) = I(\mathbf{X}; \mathbf{Y})$ for this imaging system.

\subsubsection{Direct proof of information equality}

We must show $H(\mathbf{Y} | \mathbf{O}) = H(\mathbf{Y} | \mathbf{X})$ given 
$$
I(\mathbf{O}; \mathbf{Y}) = H(\mathbf{Y}) - H(\mathbf{Y}|\mathbf{O}), \quad 
I(\mathbf{X}; \mathbf{Y}) = H(\mathbf{Y}) - H(\mathbf{Y}|\mathbf{X}).
$$
From the Markov property $\mathbf{O} \rightarrow \mathbf{X} \rightarrow \mathbf{Y}$, $\mathbf{O}$ and $\mathbf{Y}$ are conditionally independent given $\mathbf{X}$,
\begin{align}
    H(\mathbf{Y} | \mathbf{O}, \mathbf{X}) = H(\mathbf{Y} | \mathbf{X}).
\end{align}
From the deterministic condition $\mathbf{X} = f(\mathbf{O})$, knowing $\mathbf{O}$ completely determines $\mathbf{X}$, and conditioning on $\mathbf{O}$ is equivalent to conditioning on both $\mathbf{O}$ and $\mathbf{X}$, %
\begin{align}
     H(\mathbf{Y} | \mathbf{O}) = 
     H(\mathbf{Y} | f(\mathbf{O})) =
     H(\mathbf{Y} | \mathbf{O}, \mathbf{X}) 
     ).
     \label{eq:conditionalindependencefromo}
\end{align}
Therefore,
\begin{align*}
    H(\mathbf{Y} | \mathbf{O}) &= H(\mathbf{Y} | \mathbf{X}) \\
    I(\mathbf{O}; \mathbf{Y}) &= I(\mathbf{X}; \mathbf{Y}). \quad \square
\end{align*}

\subsubsection{Alternative proof of equality in Data Processing Inequality}

Since $\mathbf{O} \rightarrow \mathbf{X} \rightarrow \mathbf{Y}$ is Markov, $\mathbf{Y} \rightarrow \mathbf{X} \rightarrow \mathbf{O}$ is also Markov~\cite{dataprocessinginequality}. 

\noindent By the Data Processing Inequality (DPI)~\cite{dataprocessinginequality},
\begin{align*}
    I(\mathbf{O};\mathbf{Y}) &\leq I(\mathbf{X}; \mathbf{Y}), \\
I(\mathbf{O}; \mathbf{Y}) = I(\mathbf{X}; \mathbf{Y}) \quad &\iff \quad \mathbf{Y} \rightarrow \mathbf{O} \rightarrow \mathbf{X}.
\end{align*}
For $ \mathbf{Y} \rightarrow \mathbf{O} \rightarrow \mathbf{X}$ to be a Markov chain, we must show that $\mathbf{Y}$ and $\mathbf{X}$ are conditionally independent given $\mathbf{O}$, equivalently 
$H(\mathbf{Y} | \mathbf{X}, \mathbf{O}) = H(\mathbf{Y} | \mathbf{O})$.

As in Eq.~\eqref{eq:conditionalindependencefromo}, $H(\mathbf{Y} | \mathbf{X}, \mathbf{O}) = H(\mathbf{Y} | \mathbf{O})$. $\mathbf{Y}$ and $\mathbf{X}$ are conditionally independent given $\mathbf{O}$ and $\mathbf{Y} \rightarrow \mathbf{O} \rightarrow \mathbf{X}$ is Markov.

Therefore, by the equality condition for the DPI, 
$$I(\mathbf{O};\mathbf{Y}) = I(\mathbf{X};\mathbf{Y}) \quad \square$$

\section{Compressive 3D Imaging}
\label{sec:compressive_3D}

\begin{figure}[!t]
\centering
\includegraphics[width=0.95\columnwidth]{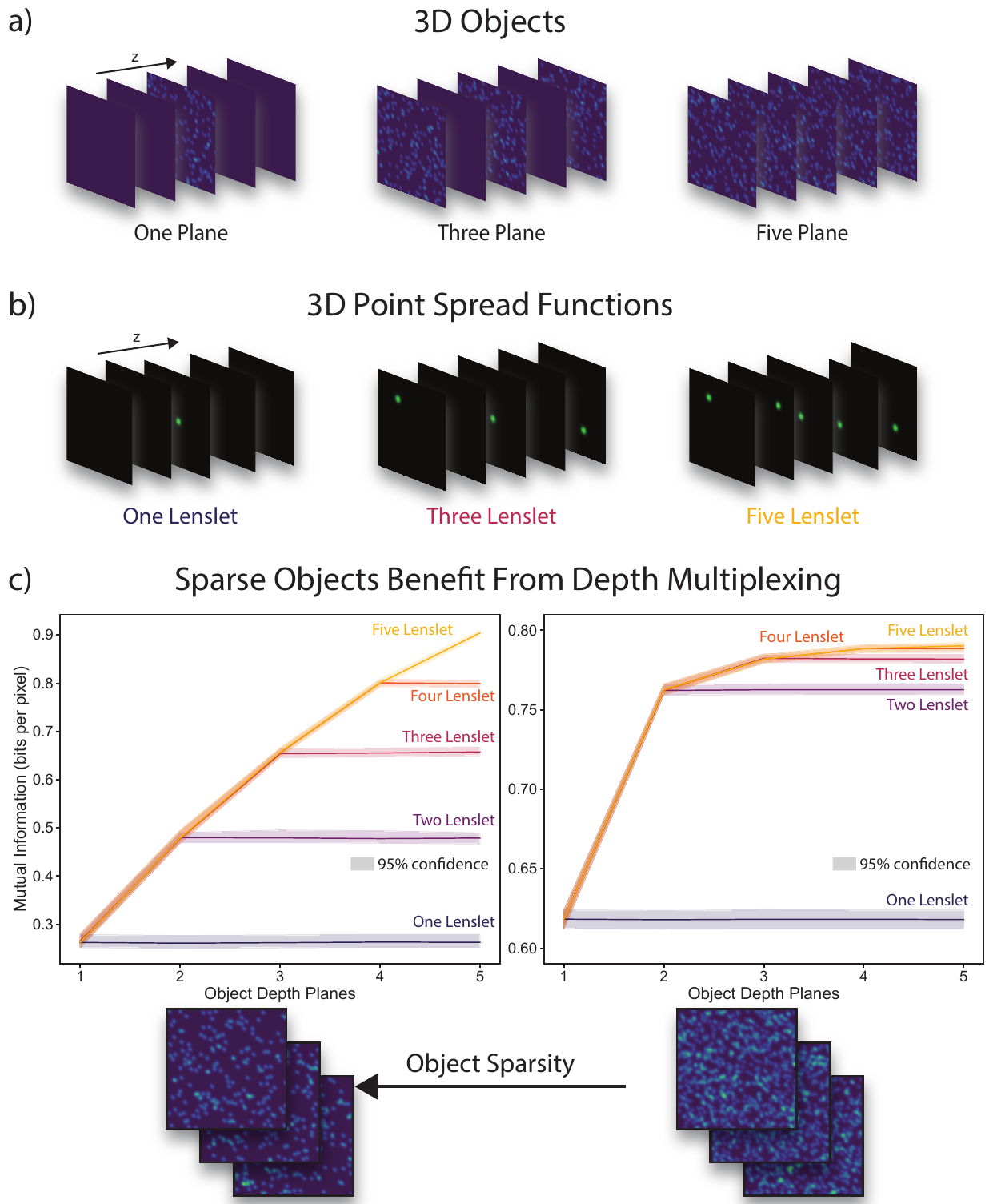}
\caption{Mutual information for depth multiplexing in single-shot 3D lensless imaging. a) Simulated 3D object distributions with one through five planes of beads. b) 3D point spread functions (PSFs) with one through five lenslets, where each lenslet is focused at a different depth plane. c) For sparse object data (Tamura Coefficient (TC) = 1.064), increasing depth multiplexing improves information. Information is maximized with the five lenslet PSF when all five depth planes are encoded in the measurement. For less sparse object data (TC = 0.783), information gain is limited with depth multiplexing. Between three and five lenslets, the information does not significantly increase.}
\label{fig:beads3D}
\end{figure}

Lensless imaging can be used for compressive imaging, as in single-shot 3D imaging, in which a 3D object is recovered from a 2D measurement that compressively encodes object depth. To explore the tradeoffs present in single-shot 3D imaging, we systematically vary objects and encoders with a similar approach to 2D imaging (Sec. 3). However, the increased dimensionality and degrees of freedom in single-shot 3D imaging opens up a complex space of parameter interactions. For simplicity, we isolate the effects of object sparsity and encoder depth multiplexing on a proof-of-concept idealized imaging system with the objective of understanding the fundamental impact of depth multiplexing.

To probe tradeoffs between 3D depth multiplexing and object sparsity, we use the bead data generation approach from Sec. 3 to simulate 3D objects. We form volumes with 1-5 planes of beads (Fig.~\ref{fig:beads3D}a), where each plane in the volume has equal object sparsity. The object depth planes are sufficiently far apart such that each bead's axial extent is contained to a single depth plane. 

For single-shot 3D imaging, the encoded image model is extended to a sum of 2D convolutions across depth planes $z$, 
\begin{equation}
\label{eq:3Dconvolution}
    \mathbf{x} = \sum_{z=1}^N \mathbf{o}_z \ast \mathbf{h}_z,
\end{equation}
where $\mathbf{o}_z$ and $\mathbf{h}_z$ are axial slices of the object and PSF, respectively. 

The encoder model is a 3D PSF consisting of lenslets with varying focal lengths, where each lenslet is focused at a different depth plane (Fig.~\ref{fig:beads3D}b). The spacing between PSF depth planes is large enough that only in-focus lenslets contribute to the measurement, with negligible contribution from defocused lenslets from other planes. This simplification isolates the effect of depth multiplexing without introducing additional complexities. In practice, this model represents an imaging system with extremely poor depth resolution, shallow depth-of-focus, and severe depth ambiguity due to minimal per-depth multiplexing. Future work can incorporate realistic features such as defocus and higher multiplexing for each depth plane to further explore optimal 3D multiplexing designs for lensless imaging systems~\cite{xueCM2, myrml, diffusercam}. 

The object distributions consist of 50,000 bead volumes with $96 \times 96 \times 5$ voxels and a mean photon count of 100. For each combination of 3D object and 3D PSF, measurements are generated based on the 3D image formation model in Eq.~\eqref{eq:3Dconvolution}. %

Figure~\ref{fig:beads3D}c summarizes the interactions between object sparsity and encoder depth multiplexing for single-shot 3D imaging. Sparse objects (left, TC = 1.064) benefit from depth multiplexing, as mutual information continues to increase with more lenslets. Information is maximized with the five lenslet encoder, which captures five object depth planes in the measurement. 
For dense objects (right, TC = 0.783), although mutual information increases with depth planes from one through three lenslets, there is negligible information gain for four and five lenslets and a lower maximum information value than for sparse objects. Intuitively, although more depth planes are incorporated by increasing depth multiplexing, the measurements for dense objects become saturated and similar, which does not provide an advantage for discriminating different objects from their corresponding measurements. This reveals an important distinction: while additional depth multiplexing may not improve information (discrimination), it is still necessary for compressive encoding, as we cannot reconstruct object features in planes that have not been measured.

\section{Sparsity and Multiplexing Tradeoffs}
\subsection{Sparsity and Multiplexing for Diffuser Encoders}
\label{sec:diffusermultiplexing}
In Sec. 3 we studied tradeoffs between object sparsity and encoder multiplexing for lenslet-based encoders. We found that sparse objects benefit from more multiplexing, and that optimally-encoded measurements achieved a constant level of measurement sparsity. Here, we verify that these findings generalize to other classes of encoders by studying diffuser-based encoders~\cite{diffusercam}.

\begin{figure}[!t]
\centering
\includegraphics[width=0.7\textwidth]{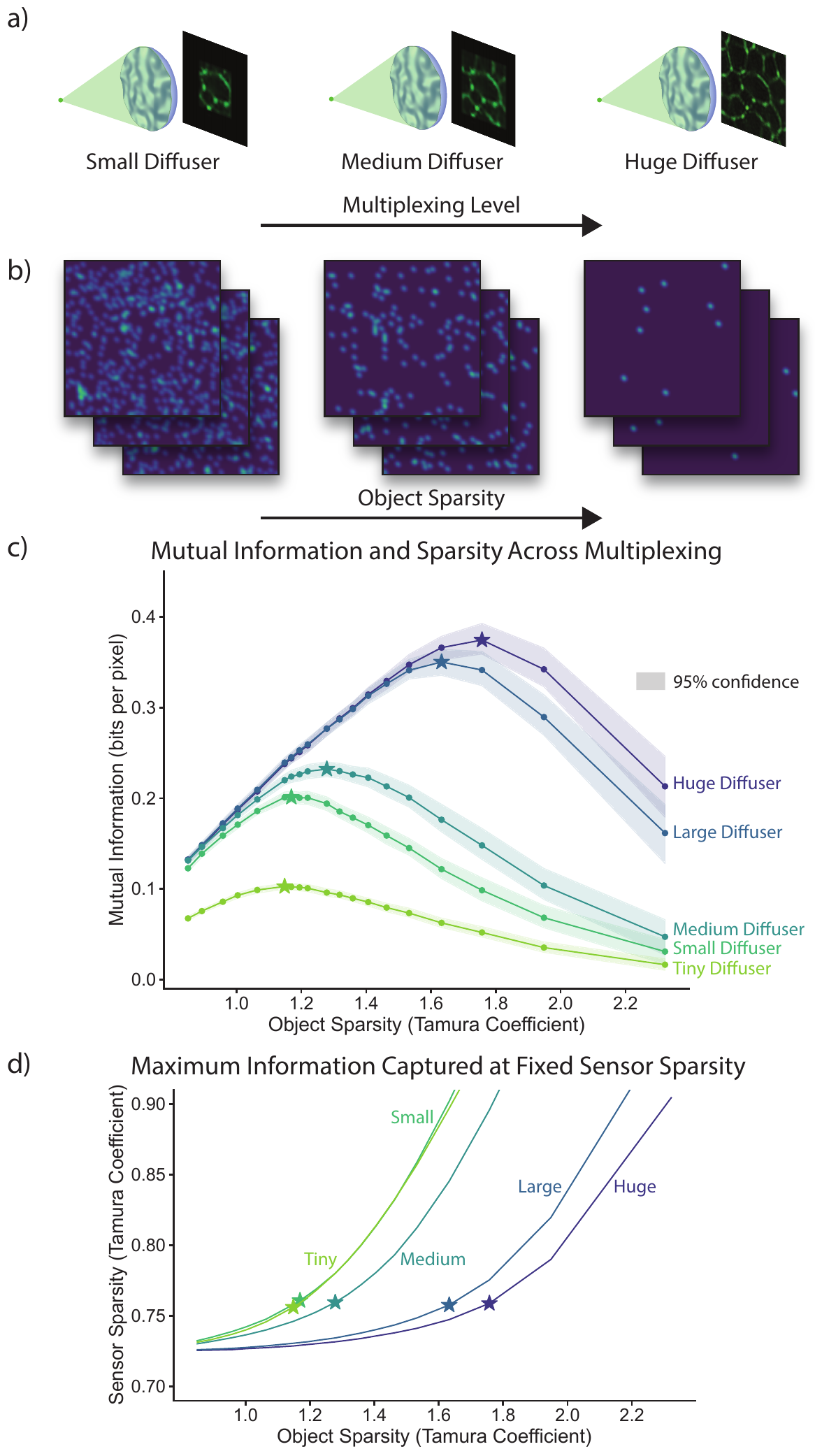}
\caption{Mutual information for varying object sparsity and encoder multiplexing for diffuser-based encoders. a) Examples of encoders with increasing levels of multiplexing, where an aperture controls the extent of the encoder point spread function (PSF). Larger encoder extent corresponds to more multiplexing. b) Examples of simulated objects with increasing levels of sparsity as quantified by the Tamura coefficient (TC). c) Mutual information for diffuser-based encoders swept across multiplexing levels. As multiplexing increases, the maximum mutual information (denoted by stars) is achieved at higher sparsity levels. d) Maximum mutual information (denoted by stars) corresponds to a fixed sensor sparsity value across all multiplexing encoders.}
\label{fig:diffusersparsity2D}
\end{figure}

Diffuser-based encoders use a Gaussian diffuser, which is a phase mask with a high-multiplexing caustic point spread function (PSF). We study five levels of multiplexing based on the spatial extent of the diffuser PSF (Fig.~\ref{fig:diffusersparsity2D}a). An aperture is used to control the spatial extent, and the highest-multiplexing encoder is the Huge Diffuser, which has no aperture. We generated datasets of objects with varying sparsity (TC = 0.850 to TC = 2.322) following the data generation approach used in Sec. 3. Example objects are visualized in Fig.~\ref{fig:diffusersparsity2D}b. Measurements were generated following the imaging pipeline in Fig. 1a.

In Fig.~\ref{fig:diffusersparsity2D}c, we study the effects of object sparsity and encoder multiplexing on mutual information. For high-multiplexing diffuser encoders, such as the Huge Diffuser, maximum information encoding corresponds to extremely high levels of object sparsity. For lower-multiplexing encoders, such as the Small Diffuser, maximum information is achieved with moderately sparse objects. The design implications are the same as for lenslet-based encoders: for maximum information encoding, highly sparse objects benefit from more multiplexing.

In Fig.~\ref{fig:diffusersparsity2D}d, we verify that all optimally-encoded measurements have the same measurement sparsity. For each multiplexing encoder, we plot the increasing relationship between object and sensor sparsity and denote the sensor sparsity corresponding to maximum mutual information with a star. Therefore, optimally-encoded measurements for diffuser-based encoders also have a constant sensor sparsity (TC $\approx$ 0.75) across multiplexing levels.

\subsection{Sparsity and Multiplexing for Read Noise-Limited Systems}
\label{sec:readnoisemultiplexing}

In Sec. 3, we studied tradeoffs between object sparsity and encoder multiplexing for shot noise-limited systems with signal-dependent Poisson noise. Here, we study read noise-limited systems with signal-independent Gaussian noise, corresponding to low-light imaging conditions.

To calculate mutual information, we follow the same procedure as in the main text and change the conditional entropy term to account for Gaussian (read) noise. Assuming that the noise at each pixel in an $M$-pixel measurement is independent of other pixels~\cite{goodmanstatisticaloptics}, the closed-form expression for the conditional entropy for Gaussian noise is
\begin{equation}
    \label{eq:MIconditionalentropygaussian}
    H(\mathbf{Y} | \mathbf{X}) = \frac{M}{2} \log_2  (2 \pi e \sigma^2), 
\end{equation}
with noise variance $\sigma^2$~\cite{Pinkard2024}.

Our study follows the same approach as in Sec. 3, using datasets of objects with varying sparsity (TC = 0.739 to TC = 1.359) and mean photon count of 100 photons, and considering multiplexing encoders ranging from a single lens to nine lenslets, as visualized in Fig. 1c. Measurements are generated following the imaging pipeline in Fig. 1a. We add signal-independent read noise with variance $\sigma^2 = 25$ to form the noisy sensor measurements $\mathbf{Y}$.

\begin{figure*}[!t]
\centering
\includegraphics[width=0.75\textwidth]{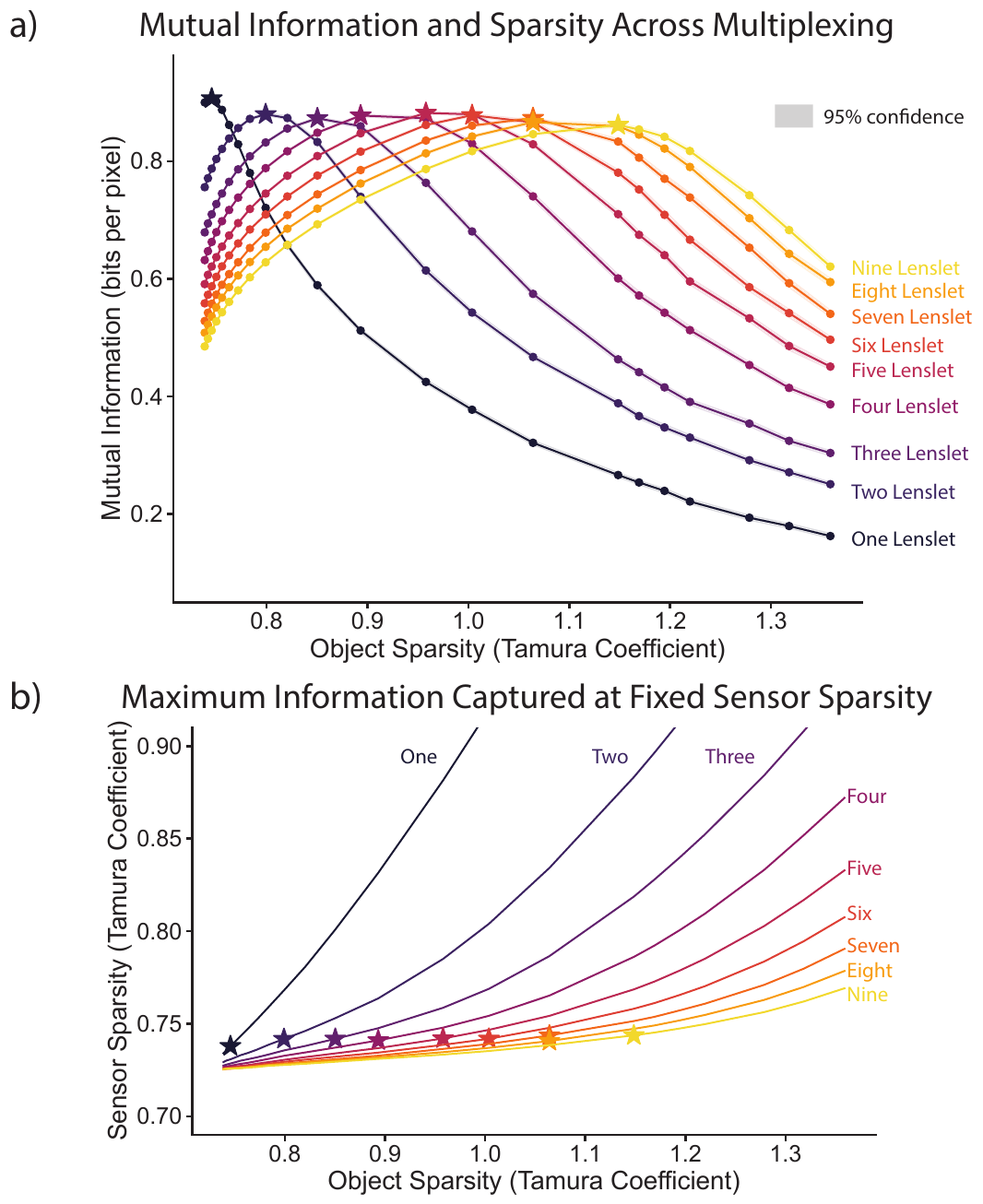}
\caption{Mutual information for varying object sparsity and encoder multiplexing under read noise with $\sigma^2 = 25$. a) Mutual information for one through nine lenslet multiplexing systems, each swept across object sparsity levels. As multiplexing increases, the maximum mutual information (denoted by stars) is achieved at higher sparsity levels. b) Maximum mutual information (denoted by stars) corresponds to a fixed sensor sparsity across all multiplexing encoders. These results are consistent with the main text experiments using shot noise.}
\label{fig:beadreadnoise2D}
\end{figure*}

In Fig.~\ref{fig:beadreadnoise2D}a, we study the tradeoff between object sparsity and encoder multiplexing. As in the shot noise case from the main text, as multiplexing levels increase, mutual information is maximized (denoted by stars) with increasingly sparse objects. 

In Fig.~\ref{fig:beadreadnoise2D}b, we verify that all optimally-encoded measurements share the same measurement sparsity. For each multiplexing encoder, we plot the relationship between object and corresponding sensor sparsity, and label the sensor sparsity corresponding to maximum mutual information with a star. Optimally-encoded measurements with read noise have a constant sensor sparsity of TC $\approx$ 0.74 across multiplexing levels, which is very close to the value found for shot noise (TC $\approx$ 0.75). 

\subsection{Sparsity and Multiplexing with Maximum Value Object Normalization}
\label{sec:maxnormmultiplexing}

In Sec. 3, we studied tradeoffs between object sparsity and encoder multiplexing for shot noise-limited systems and objects normalized to have a constant mean photon count. Here, we study shot noise-limited systems using objects normalized to have constant maximum photon count.

\begin{figure*}[!t]
\centering
\includegraphics[width=0.65\textwidth]{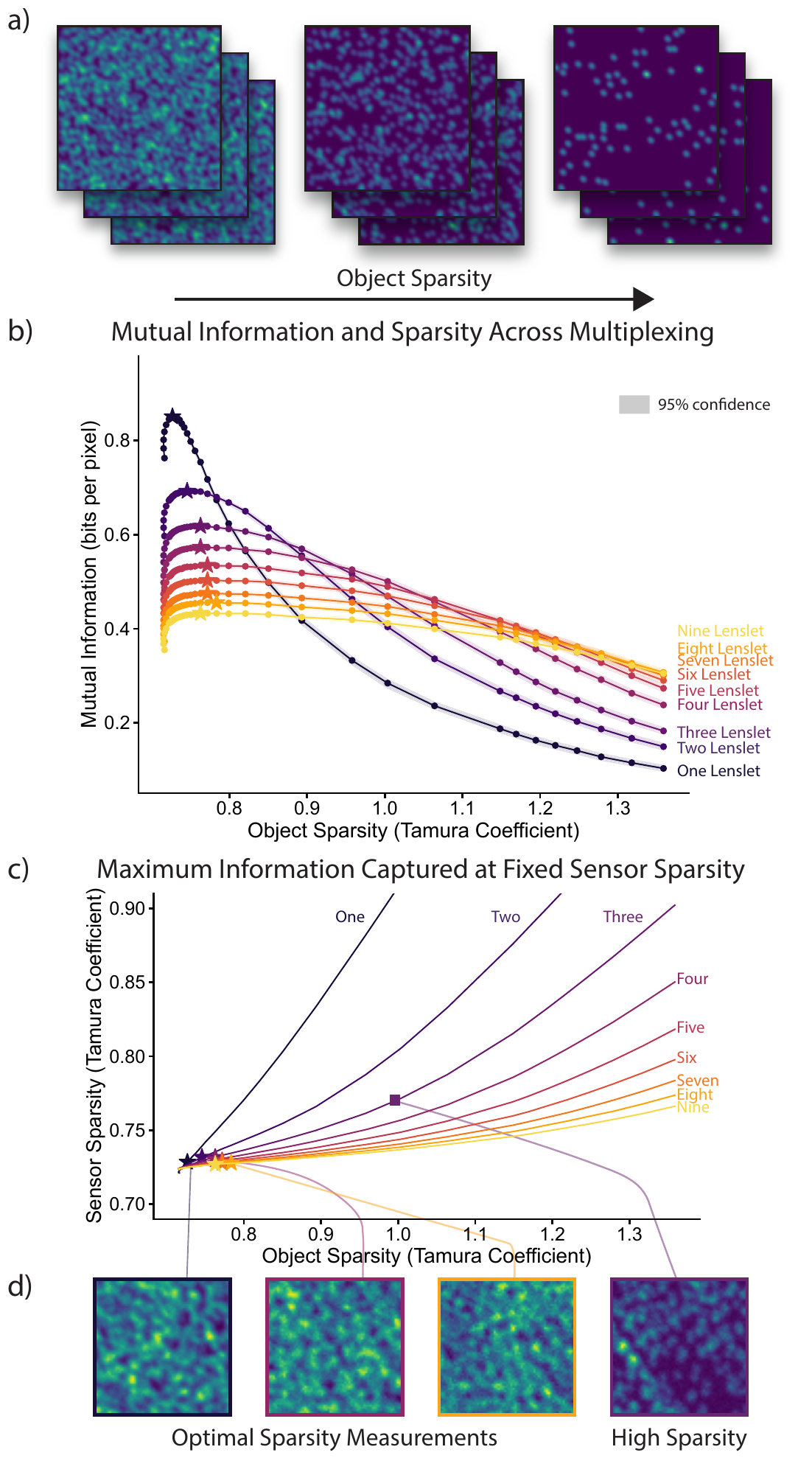}
\caption{Mutual information for varying object sparsity and encoder multiplexing with maximum per-bead normalization. a) Examples of simulated objects with increasing levels of sparsity as quantified by the Tamura coefficient. b) Mutual information for one through nine lenslet multiplexing systems, each swept across object sparsity levels. As multiplexing increases, the maximum mutual information (denoted by stars) is achieved at higher sparsity levels. c) Maximum mutual information (denoted by stars) corresponds to a fixed sensor sparsity across all multiplexing encoders. d) Example measurements corresponding to fixed sensor sparsity appear similar, and differ from those with non-fixed sparsity.}
\label{fig:beadmaxnorm2D}
\end{figure*}

Our study follows the same approach as in Sec. 3, using datasets of objects with varying sparsity (TC = 0.716 to TC = 1.359), and considering multiplexing encoders ranging from a single lens to nine lenslets. We assign a maximum photon count of 1000 photons per bead, corresponding to constant photon emission per bead for maximum value normalization (Fig.~\ref{fig:beadmaxnorm2D}a).

In Fig.~\ref{fig:beadmaxnorm2D}b, we study the tradeoff between object sparsity and encoder multiplexing. As in the main text, as multiplexing levels increase, mutual information is maximized (denoted by stars) with increasingly sparse objects. In Fig.~\ref{fig:beadmaxnorm2D}c, we verify that all optimally-encoded measurements share the same measurement sparsity by plotting the relationship between object and corresponding sensor sparsity for each multiplexing encoder, labeling the sensor sparsity corresponding to maximum mutual information with a star. Optimally-encoded measurements with maximum value object normalization have a constant sensor sparsity of TC $\approx$ 0.73 across multiplexing levels. Examples of measurements with constant sparsity in Fig.~\ref{fig:beadmaxnorm2D}d are qualitatively similar and differ from an example with non-optimal sparsity. Although maximum value normalization yields different absolute mutual information values and curve shapes, our key finding of constant measurement sparsity remains consistent with mean normalization-based evaluation. This robustness confirms that this sparsity principle reflects fundamental encoding properties rather than normalization effects.

\section{Experimental Lensless Imaging}

\begin{figure*}[!t]
\centering
\includegraphics[width=\textwidth]{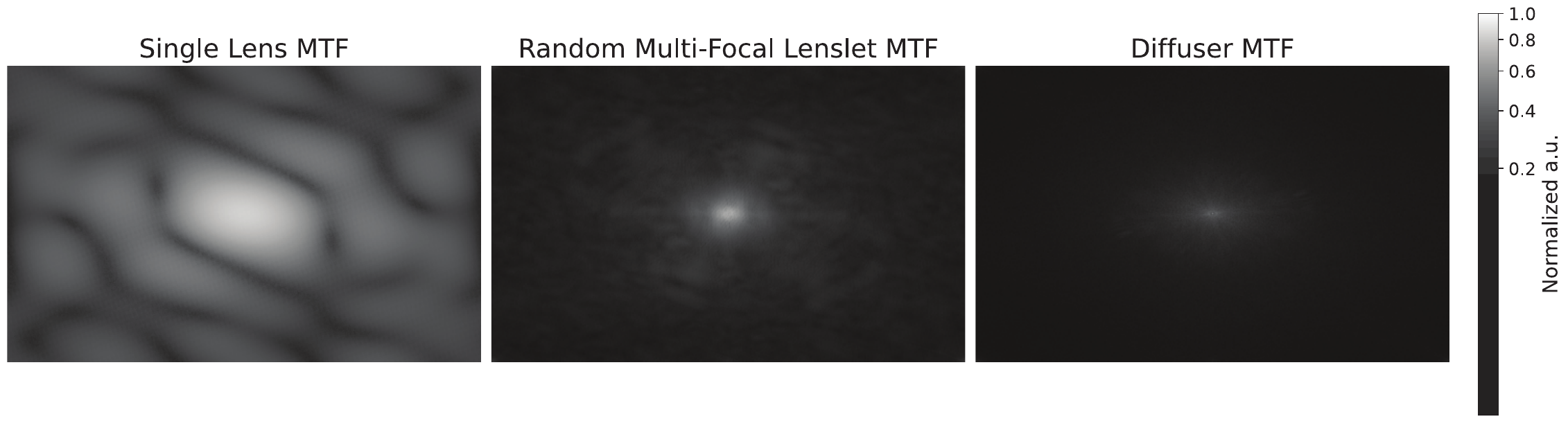}
\caption{Visualizations of modulation transfer functions (MTFs) for the experimentally-captured point spread functions from single lens, random multi-focal lenslet (RML), and diffuser imaging systems.}
\label{fig:2Dmtfs}
\end{figure*}

In the main text (Fig. 4), we plot each imaging system PSF's modulation transfer function (MTF). Specifically, a 1D cross-section horizontally through the DC component. These MTFs are not radially symmetric. Therefore, we do not report radially-averaged MTFs. Instead, we also visualize the complete 2D MTF for each imaging system in Fig.~\ref{fig:2Dmtfs}.

To reconstruct images from measurements from the diffuser and RML lensless imaging systems, we use a ConvNeXt architecture~\cite{convnext}. We use experimentally-captured datasets of measurements~\cite{claradataset} with the lensed camera images serving as ground truth. For each lensless imager, the reconstruction algorithm is trained on $8 \times$ downsampled images, at a resolution of $150 \times 240$ pixels. We use a set of 24,000 measurements and a held-out test set of 1,000 measurements, and train until convergence ($< 35$ epochs) with default parameters. Reported reconstruction metrics are averaged across the held-out test set. The main text reports the Structural Similarity Index Measure (SSIM), and additional error and quality metrics are reported in Table~\ref{table:metrics}. For all metrics, the RML lensless imager outperforms the diffuser. Reconstruction performance is in agreement with the mutual information estimates in the main text.

\begin{table*}[!t]

\renewcommand{\arraystretch}{1.0}
\caption{Reconstruction Quality Metrics for Diffuser and Random Multi-Focal Lenslet (RML) Lensless Imaging Systems}
\centering
\resizebox{\textwidth}{!}{
\begin{tabular}{|c||c|c|c|c|}
\hline
& \textbf{Mean Squared Error} & \textbf{Peak Signal-to-Noise Ratio} & \textbf{Structural Similarity Index Measure} & \textbf{Learned Perceptual Image} \\
& \textbf{(MSE) $\downarrow$} & \textbf{(PSNR) $\uparrow$} & \textbf{(SSIM) $\uparrow$} & \textbf{Patch Similarity (LPIPS) $\downarrow$} \\
\hline\hline
\multirow{1}{*}{\textbf{Diffuser}} & 1.740e-4 & 28.565 & 0.821 & 0.0213 \\
\hline
\multirow{1}{*}{\textbf{RML}} & 1.259e-4 & 29.970 & 0.886 & 0.0127 \\
\hline
\end{tabular}
}
\label{table:metrics}
\end{table*}

\section{Information-optimal encoder design}
\label{supp:ideal}

\begin{figure}[!t]
\centering
\includegraphics[width=\columnwidth]{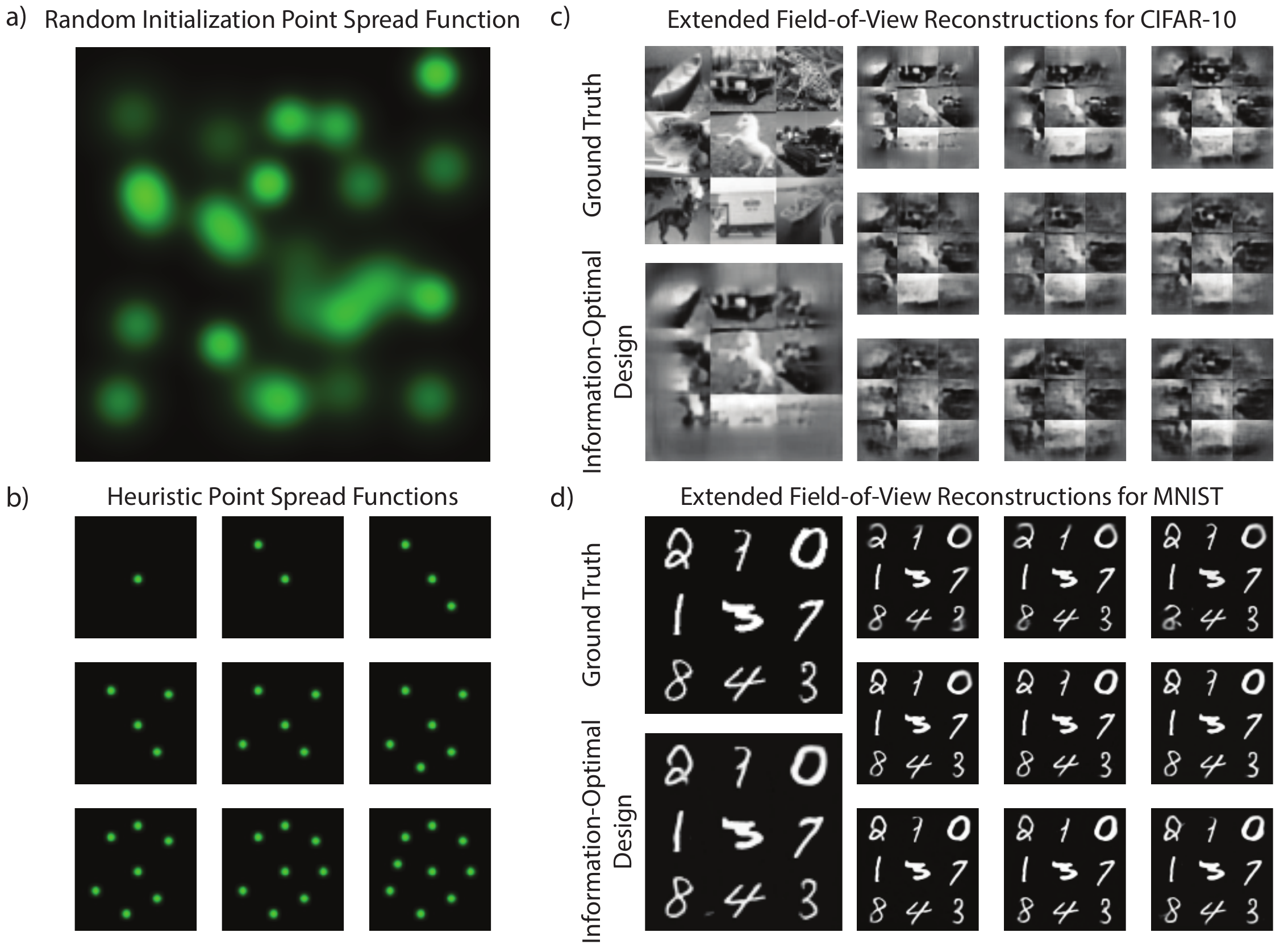}
\caption{Visualizations of point spread functions (PSFs) and object reconstructions. a) 25 lenslet PSF used for random initialization for information-optimal encoder design. b) One through nine lenslet PSFs used in all heuristic evaluations and heuristic initialization for information-optimal encoder design. c) Extended field-of-view reconstructions for one through nine lenslet PSFs and information-optimal PSF for CIFAR-10 natural objects. d) Extended field-of-view reconstructions for one through nine lenslet PSFs and information-optimal PSF for MNIST sparse objects.}
\label{fig:initialization}
\end{figure}

\subsection{Random Initialization} 
\label{supp:idealrandominit}

To account for variability due to initialization, we initialize IDEAL with 10 different random initializations and optimize for the MNIST dataset. We focus on this sparse dataset because the resulting encoder designs have the highest multiplexing, corresponding to many degrees of freedom. Each optimization starts with a different random distribution of 25 lenslets with arbitrary position, width, and weighting. The resulting IDEAL designs achieve mutual information values ranging from 0.621 to 0.666, corresponding to designs with 6 through 8 lenslets. The highest information values are achieved with 8 lenslets and the lowest achieved with 6 lenslet designs. All designs outperform the baseline heuristic designs, and the best result, the 8 lenslet design achieving 0.666 bits per pixel of information, is reported in the main text.

\subsection{Heuristic Initialization}

\begin{figure*}[]
\centering
\includegraphics[width=\textwidth]{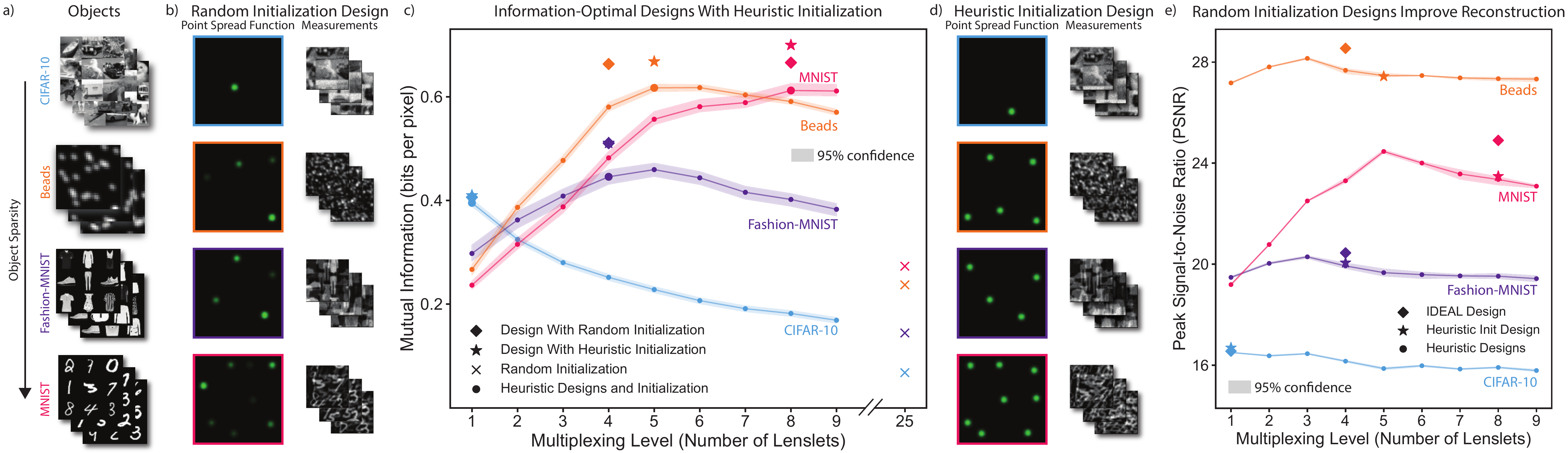}
\caption{Information-optimal phase mask design for lenslet-based encoders with Information-Driven Encoder Analysis Learning (IDEAL). 
a) Examples of objects with varying sparsity. 
b) Visualization of the encoder point spread function (PSF) and example measurements for the IDEAL design for each object distribution with a random initialization of lenslets. As the object sparsity increases, the number of lenslets and the resulting encoder multiplexing increases.
c) Information-optimal designs with random and heuristic initialization for objects with varying sparsity. Random initialization with 25 lenslets is denoted by an "X", and the resulting IDEAL design is denoted by a diamond. Mutual information for heuristic designs with one through nine lenslets are plotted for reference, with 95\% confidence intervals indicated by shading. IDEAL designs starting with a heuristic initialization are denoted by a star. Confidence intervals are smaller than the marker size for IDEAL designs. IDEAL designs have more information than heuristic designs, and IDEAL designs with heuristic initialization have slightly more information than IDEAL designs with random initialization. 
d) Visualization of the encoder PSF and example measurements for IDEAL designs with heuristic initialization of lenslets for each object distribution. These PSFs have equally-weighted lenslets and produce measurements with denser sensor coverage than in b). Optimal sensor sparsity is achieved for Beads and MNIST datasets. 
e) IDEAL designs have higher information and better reconstruction performance than the heuristic designs, as evaluated by reconstruction peak signal-to-noise ratio (PSNR). Although IDEAL designs with heuristic initialization have more information than those with random initialization, the additional information is not beneficial for compressive extended field-of-view reconstruction with a U-net.}
\label{fig:supplementIDEAL}
\end{figure*}

Heuristic lenslet designs used in the main text are visualized in Fig.~\ref{fig:initialization}b. The lenslet positions were selected to keep the lenslet distribution approximately uniform across the PSF extent. Lenslet positions remain fixed, with each increase in multiplexing adding one new lenslet position.

We also consider IDEAL with heuristic initialization to determine whether designs achieve local or global optima. We initialize IDEAL with the top three heuristic designs (by information) for each dataset. Keeping all design degrees of freedom (position, width, and weight), resulted in similar designs and information estimates as random initialization. Instead, we constrain lenslets to be equally weighted, learning only positions and widths. These constrained designs had higher information than those from random initialization.
We report the resulting design with most information for each dataset in Fig.~\ref{fig:supplementIDEAL}c, with the corresponding heuristic initialization indicated by a large circle. Heuristic initialization for IDEAL provides significant information gain for sparse objects (MNIST), but not for dense objects. By constraining our optimization problem's degrees of freedom and carefully initializing, we can further improve encoder designs and navigate the non-convex loss landscape~\cite{wolfgange2e}.

In Fig.~\ref{fig:supplementIDEAL}d, we visualize the resulting encoder PSF and example measurements for each of the heuristic initialization IDEAL designs. In the main text (Fig. 2c), we found that information-optimal systems achieved a constant sensor sparsity. Although random initialization designs do not produce measurements with this sparsity, the heuristic initialization design measurements approach this constant, and the Beads and MNIST designs exactly achieve it. For Fashion-MNIST and CIFAR-10, as object sparsity cannot be adjusted, a discrete number of lenslets is unable to exactly achieve this value. Regardless, evaluating sensor sparsity can help determine whether a design is close to the global information optimum. 

In Fig.~\ref{fig:supplementIDEAL}e, we compare reconstruction performance for heuristic designs, random initialization IDEAL designs, and heuristic initialization IDEAL designs for compressive extended-field-of-view 2D imaging. We reconstruct objects from multiplexed measurements using a lightweight U-net (Supplement Sec.~\ref{supp:implementation}), and evaluate reconstruction performance based on the average test set PSNR, including 95\% confidence intervals. In general, both types of IDEAL designs outperform the heuristic baseline in reconstruction performance for each object dataset. Interestingly, although heuristic initialization IDEAL designs encode more information than random initialization IDEAL designs, their reconstruction performance is generally worse than for random initialization IDEAL designs. This highlights a subtlety: not all encoded information can be effectively utilized by every reconstruction algorithm, especially those with limited power and flexibility. Even in general tasks like reconstruction, some of the encoded information may not be relevant. In addition, error metrics such as PSNR may fail to fully capture the benefits of those bits of information. Future work can investigate more powerful reconstruction architectures, improved reconstruction metrics, and task-specific conditioning of mutual information. Regardless, maximizing encoded information offers clear performance gains and useful design principles.

\subsection{Lenslet Aperture Shape} 
In information-optimal design, we learn isotropic covariance matrices, which constrain lenslet apertures to be round. We additionally designed lenslets with arbitrary aperture shapes, represented by anisotropic covariance matrices. The resulting design for MNIST was suboptimal, converging to 6 lenslets with mutual information of 0.552 bits per pixel.

\end{document}